
\input amstex
\magnification 1200
\documentstyle{amsppt}
\NoBlackBoxes
\NoRunningHeads
\def\C{\Bbb C}

\def\Z{\Bbb Z}
\def\v{^\vee}

\def\d{\partial}

\def\g{\frak g}
\def\sln{\frak{sl}_n}

\def\Tr{\text{\rm Tr}}

\def\l{\lambda}
\def\lhat{{\hat \lambda}}
\def\rhat{{\hat\rho}}
\def\dhat{{\hat\delta}}
\def\ha{\hat\alpha}
\def\a{\alpha}
\def\<{\langle}
\def\>{\rangle}
\def\CK{\C[\hat P_k]^{\hat W}}

\topmatter
\title Spherical functions on affine Lie groups
\endtitle
\author {\rm {\bf Pavel I. Etingof, Igor B.Frenkel,
 Alexander A. Kirillov, Jr.} \linebreak
\vskip .1in
   Department of Mathematics\linebreak
   Yale University\linebreak
   New Haven, CT 06520, USA\linebreak
   e-mail: etingof\@math.yale.edu,
			kirillov\@math.yale.edu}
\endauthor
\endtopmatter
\document

\centerline{July 7, 1994}
\vskip .05in
\centerline{hep-th 9407047}
\vskip .05in
\centerline{\bf Introduction}
\vskip .05in

By spherical functions one usually means functions
on the double coset space $K\backslash G/K$, where $G$ is a group and
$K$ is a subgroup of $G$. This is equivalent to considering
functions on the homogeneous space $G/K$ left invariant with respect
to $K$. More generally, if $V$ is a fixed irreducible representation of $K$,
for example, finite-dimensional, one can look at functions on $G/K$
whose left shifts by elements of $K$ span a space which is isomorphic
to $V$ as a $K$-module. Consideration of such functions is equivalent to
consideration of functions on $G/K$ with values in the dual representation
 $V^*$ which are equivariant with respect to the left action of $K$.
In this (and even more general) framework
spherical functions were studied in the works of
Harish-Chandra, Helgason and other authors \cite{HC,He,W}.

In the classical theory of spherical functions, $G$ is often a
real non-compact Lie group, and $K$ is a maximal compact subgroup of
$G$. In this case, $G/K$ is a non-compact symmetric space.
One can also consider an associated compact symmetric space $G_c/K$,
where $G_c$ is a compact form of $G$. An important class of examples
is complex semisimple groups considered as real groups.
In this case $G_c=K\times K$, and $K$ is embedded diagonally into $G$.
The study of $K$-equivariant functions on $G/K$ is then
equivalent to the study of functions on $K$ itself equivariant
with respect to conjugacy. This problem makes sense for an arbitrary
group $K$, and it turns out that equivariant functions can be
explicitly constructed as traces
of certain intertwining operators. In this paper we describe such
functions in two cases -- $K$ is a compact simple Lie group, and $K$ is
an affine Lie group (i.e. an infinite-dimensional group
whose Lie algebra is an affine Lie algebra).

The results concerning the compact group case are given in Chapter 1.
For a compact Lie group $K$ and a pair $W,V$ of irreducible finite-dimensional
representations of $K$ we consider an intertwining
operator $\Phi: W\to W\otimes V^*$ and associate with it the
 function $\Psi(x)=\Tr|_W(\Phi x)$. This function takes values
in $V^*$ and is equivariant with respect to conjugacy, and the Peter-Weyl
theorem implies that all equivariant functions can be written
as linear combinations of such functions.

The next step is computation of the radial parts of the Laplace
operators of $K$ acting on conjugacy equivariant functions.
This means, rewriting these operators in terms of the coordinates on the set
of conjugacy classes, which is the maximal torus in $K$. The result
is a completely integrable quantum system with matrix coefficients --
a set of $r$ commuting differential operators in $r$ variables
symmetric with respect to the Weyl group, where $r$ is the rank of
$K$. In a special case, this system coincides with the
trigonometric Calogero-Sutherland-Moser multi-particle system
(\cite{C,S,OP}). This allows to express eigenfunctions of this system,
in particular, Jack's symmetric functions, in terms of traces of
intertwining operators \cite{E,EK2,ES}.

These results basically fit into the framework of the classical theory
of spherical functions on symmetric spaces,
and are formulated mostly for the purpose of motivation.
 Most of them can be deduced from
the classical results of Harish-Chandra and Helgason
\cite{HC,He}.

The results concerning the affine case are given in Chapter 2.
We study holomorphic functions on a complex affine Lie group
taking values in an irreducible finite-dimensional representation
of this group and equivariant with respect to (twisted) conjugacy.
We prove that the space of equivariant functions
having a fixed homogeneity degree with respect to the action
of the center of the group
is finite-dimensional. We show that this space can be
nontrivial only for positive integer values of the degree, and in this case
it coincides with a certain space of intertwining
operators between representations of the affine Lie algebra, (cf.
\cite{TK},\cite{FR}). This is done by constructing a basis of the
space of equivariant
functions consisting of
weighted traces of the intertwining operators. These functions are
affine analogues of the functions $\Psi(x)$ defined above for the
compact group.

We compute the radial part of the second order Laplace
operator on the affine Lie group acting on equivariant
functions, and find that it is a certain
 parabolic partial differential operator.
Weighted traces of intertwiners form an eigenbasis for this
operator in the space of equivariant functions.
In a special case, they coincide (up to a certain factor)
with the affine Jack's polynomials which were defined in \cite{EK3}.

At the critical value
of the degree (minus the dual Coxeter number of the underlying
simple Lie algebra) there exist higher order Laplace operators
coming from the corresponding central elements of the completed
universal enveloping algebra of the affine Lie algebra
(\cite{Ma,H,GW2,FF}). All these
operators commute with each other, and their radial parts form a
commutative system of differential operators with elliptic coefficients.
In a special case, this system coincides with the elliptic
Calogero-Sutherland-Moser multi-particle system (\cite{OP}).
In particular, for the affine Lie group $\widetilde{SL}_2$,
we obtain the classical Lame operator \cite{WW}. This shows
that the classical Lame polynomials (eigenfunctions of the Lame
operator) ``live'' on the group $\widetilde{SL}_2$,
in the same sense as Legendre and Gegenbauer polynomials
``live'' on the group $SU(2)$.

When the elliptic modulus $t$ goes to $\infty$
(the elliptic curve degenerates to the rational curve),
the theory of spherical functions on the affine group
degenerates to the theory of spherical functions on the corresponding
compact group.  At the level of formulas,
this reduces to replacing elliptic functions by their trigonometric
limits. In particular, instead of the elliptic version
of the Calogero-Sutherland-Moser system one obtains its
trigonometric version.

In Chapter 3 we explain the connection between conjugacy classes
of an affine Lie group and holomorphic principal bundles,
and give proofs of the results of Chapter 2.

We would like to emphasize that in the case of an affine Lie group
classical notions of the theory of spherical functions
and Laplace operators correspond to
important objects in quantum field theory.
In particular, our results show that the space of equivariant functions
on an affine Lie group of a fixed degree coincides with
the space of conformal blocks of the Wess-Zumino-Witten
conformal field theory on an elliptic curve
with one or several punctures (cf. \cite{MS}), or,
equivalently, with the space of states of the Chern-Simons
topological field theory in genus 1 \cite{FG}. This provides a group-theoretic
realization of the modular functor for elliptic curves (cf. \cite{Se}).
Also, it turns out that the
the radial part of the second order Laplace operator
on an affine Lie group acting in the space of equivariant
functions coincides with the operator defining the Knizhnik-Zamolodchikov
connection on conformal blocks on elliptic curves, and its
eigenfunctions coincide with the
correlation functions of conformal blocks (cf. \cite{Be},\cite{FG},\cite{EK1}).

It is worth mentioning that
the computation of the radial part of Laplace operators
for the compact and affine case is the quantum analogue of
the infinite-dimensional Hamiltonian reduction
procedure which allows to obtain classical finite-dimensional
integrable systems of Calogero type from infinite-dimensional systems.
This procedure was first described in \cite{KKS} for the simple group
and simple Lie algebra case, and later generalized
to the loop and double loop case in \cite{GN1},\cite{GN2}.

Finally, we would like to remark that the results of this paper
can be generalized to the case of quantum groups and quantum
affine algebras. This amounts to a systematic study of traces
of intertwining operators for these algebras.
Results in this direction, connecting the
traces with Macdonald's symmetric functions and their generalizations,
were obtained in \cite{EK3,EK4}. The recently introduced notion
of a quantum symmetric space (see, e.g. \cite{N}, and references therein)
allows to give a ``q-geometric'' interpretation of these results, i.e.
in terms of the theory of ``q-spherical functions''. This
is a subject of future research.

\vskip .1in
\centerline {\bf Acknowledgements}
\vskip .1in

The authors would like to thank K.Gawedzki, I.Grojnowski,
Yu.Neretin and G.Zuckerman for useful
discussions. The authors are grateful to G.Felder for correcting
some errors in the first version of this paper.

The work of P.E. was supported by Alfred P.Sloan graduate dissertation
fellowship. The work of I.F. was supported by the NSF grant DMS-9400908.

\vskip .1in
\centerline{\bf 1. Spherical functions}
\vskip .1in

{\bf 1.1. Spherical functions on groups.}

Let $K$ be a group. Let $K_d$ be the
diagonal subgroup in $K\times K$: $K_d=\{(k,k)\in K\times K |k\in
K\}$. Consider the symmetric space $X_K=K\times K/K_d$. It is isomorphic
to $K$ as a set: the isomorphism is given by $(k,l)\to kl^{-1}$.
Under this map, the left action of $K\times K$
on $X_K$ is transformed into the two-sided action of $K$ on itself,
and the action of $K_d$ on $X_K$ becomes the action of $K$ on itself
by conjugacy.

\proclaim{Definition 1.1A} A function $f:X_K\to\C$ is called
spherical if the vector space $V_f$ spanned by the functions
$f^g(x)=f(g^{-1}x)$, $x\in X_K$, for all $g\in K_d$, is
finite-dimensional.
\endproclaim

Since $X_K$ is isomorphic to $K$ itself, this definition
is equivalent to

\proclaim{Definition 1.1B} A function $f:K\to \C$ is called
spherical if the vector space $V_f$ spanned by the functions
$f^g(y)=f(g^{-1}yg)$, $y\in K$, for all $g\in K$, is
finite-dimensional.
\endproclaim

We will use Definition 1.1B unless otherwise specified.

If $f$ is a spherical function then the space $V_f$ is naturally a
representation of $K$.

\proclaim{Definition 1.2} Let $V$ be a finite-dimensional
representation of $K$. A spherical function $f$ is of type $V$ if
$V_f$ is isomorphic to $V$. In particular, $f$ is called a central
function (or a class function) if
$V_f=\C$ is the trivial representation.
\endproclaim

Suppose that $f:X_K\to \C$ is a spherical function of type $V$.
We assume that $V_f$ is somehow identified with $V$. Let $f_1,...,f_n$ be a
basis of $V_f$, and let $f^1,...,f^n$ be the dual basis of $V_f^*$.
We can regard $f^1,...,f^n$ as vectors in $V^*$.
Define a vector-valued function $\Psi_f(x)\in V^*$:
$$
\Psi_f(x)=\sum_{i=1}^nf_i(x)f^i.\tag 1.1
$$
It is clear that $\Psi_f$ is determined uniquely by the
space of functions $V_f$ and its identification with $V$.

\proclaim{Definition 1.3} Let $V$ be a
representation of $K$. A function $\Psi: K\to V^*$ is called
(conjugacy) equivariant if
$g^{-1}\Psi(gxg^{-1})=\Psi(x)$,
$g\in K$.
\endproclaim

The following elementary proposition describes
the properties of $\Psi_f$ and the connection between
$f$ and $\Psi_f$.

\proclaim{Proposition 1.1}

(i) $\Psi_f$ is equivariant.

(ii) Let $V$ be an irreducible representation of $K$, and
$\Psi: K\to V^*$ be a nonzero equivariant function.
Then for any nonzero
vector $v\in V$ the function $f(x)=\langle v,\Psi(x)\rangle $
is spherical of type $V$.

(iii) Any spherical function of type $V$
can be written as $\langle v,\Psi(x)\rangle $, where $v\in V$ and $\Psi$ is
equivariant ($\Psi$ can be chosen to be $\Psi_f$).
If $V$ is irreducible then
$v,\Psi(x)$ are determined uniquely
by $f$ up to a factor; in particular, $\Psi(x)=\Psi_f(x)$ for
a suitable identification $V_f\to V$.

(iv) If every finite-dimensional
representation of $K$ is semisimple (e.g. $K$ is a finite group or a
compact Lie group)
then any spherical function is a sum of spherical functions
of irreducible types.
\endproclaim

This proposition shows that in order to understand spherical functions
on $K$ (with respect to conjugacy), it is enough to understand
equivariant functions $\Psi: K\to V^*$, where $V$ is
finite-dimensional. Therefore, from now on
we will mostly work with equivariant functions.

A typical example of a conjugacy invariant function on $K$
is a character of an irreducible finite-dimensional
representation of $K$. Similarly, an example of a vector-valued
conjugacy equivariant function is a ``vector-valued character''.
Namely, let $V,W$ be finite-dimensional representations
of $K$. Let $\Phi: W\to W\otimes V^*$ be an intertwining operator.
Consider the function $\Psi:K\to V^*$ given by
$$
\Psi(x)=\Tr|_{W}(\Phi x).\tag 1.2
$$
It is easy to see that this function is equivariant.
If $W$ is irreducible and $V=\C$ then $\Phi$ is proportional to the identity
and thus $\Psi$ is proportional to the character of $W$.
In general it is natural to call $\Psi$ a vector-valued character.

\vskip .1in

{\bf 1.2. Spherical functions on finite and compact groups.}

Let $K$ be a finite group or a compact Lie group.
Then the characters of irreducible
representations of $K$ are pairwise orthogonal and span
the space of conjugacy invariant functions. A similar statement
holds for vector-valued characters. Namely, let $W$ be
an irreducible representation of $K$, and let $S(V,W)$ be the space
of functions spanned by the functions (1.2) for all choices of $\Phi$.
The correspondence $\Phi\to f$ given by (1.2) is a bijection,
thus the space $S(V,W)$ is isomorphic to $\text{Hom}_K(W\otimes W^*,V^*)$.

Let $K^{\wedge}$ denote the set of irreducible finite-dimensional
representations of K.

Since $V$ is a finite-dimensional representation, it is unitarizable,
so there exists a natural Hermitian inner product $(,)$ on $L^2$-functions
from $K$ to $V^*$.

\proclaim{Proposition 1.2}

(i) (Weyl orthogonality)
The spaces $S(V,W)$ are pairwise orthogonal with respect to $(,)$
for $W\in K^{\wedge}$.

(ii) (Maschke-Peter-Weyl theorem)
Let $F^V(K)$ be the space of all conjugacy equivariant
$L^2$-functions $\Psi: K\to V^*$. Then $F^V(K)=\oplus_{W\in
K^{\wedge}}S(V,W)$. Thus, $F^V(K)$ is isomorphic to
$\text{Hom}_K(H,V^*)$, where
$H=\oplus_{W\in K^{\wedge}}W\otimes W^*$
(in the Lie group case the direct sum and Hom should be understood in
in $L^2$-sense).
\endproclaim

\vskip .1in
{\bf 1.3. Laplace operators on a compact Lie group.}

Now let $K$ be a compact simply connected simple Lie group.
Let us consider Laplace operators on $K$.

Let $\frak k$ be the Lie
algebra of $K$, and let $\frak g$ be its complexification.
Let $\frak h$ be a Cartan subalgebra in $\frak g$.
We identify $\frak h$ and $\frak h^*$ by means of the invariant form
on $\frak g$ in which long roots satisfy $\langle \alpha,\alpha\rangle =2$.

To every element $Y$ in the universal enveloping
algebra $U(\frak g)$ there corresponds a left-invariant
differential operator $D_Y$ on $K$ (with complex coefficients).
This operator is also right invariant iff $Y\in\Cal Z(\frak g)$,
where $\Cal Z(\frak g)$ is the center of $U(\frak g)$.
Thus, to any element $Y\in \Cal Z(\frak g)$ there corresponds
a differential operator $D_Y:F^V(K)\to F^V(K)$ for any $V$, and
$[D_{Y_1},D_{Y_2}]=0$ for $Y_1,Y_2\in\Cal Z(\frak g)$.
These operators satisfy the following obvious property:

\proclaim{Proposition 1.3} The spaces $S(V,W)\subset F^V(K)$ are eigenspaces
of $D_Y$ with eigenvalues $Y|_W=\chi(Y)(\lambda_W)$, where
$\chi: \Cal Z(\frak g)\to U(\frak h)$ is the Harish-Chandra
homomorphism,
and $\l_W\in\frak h^*$ is the highest weight of $W$.
\endproclaim

This is a vector-valued version of the well known fact
that characters of irreducible representations of $K$ are
eigenfunctions of the Laplace operators.

\vskip .1in

{\bf 1.4. Radial parts of Laplace operators.}

A beautiful application of this construction is the theory
of radial part of Laplace operators. The main idea is that
an equivariant function is uniquely defined by its
restriction to the maximal torus $T\subset K$, since any element of
$K$ can be brought into $T$ by a conjugation.
Moreover, the restriction of an equivariant function to $T$
has to take values in the zero weight space. Therefore,
the differential operator $D_Y$ can be written in terms of
the coordinates along $T$, and the resulting expression is a
differential operator with coefficients in $\text{End}V^*[0]$.
We call this expression the radial part of $D_Y$ on $F^V(K)$ and
denote it $R_V(Y)$.

Let $R$ be the root system of $\frak g$,
$W$ be the Weyl group,
$R^+$ be the set of positive roots. Let $Q^+$ be the semigroup
generated by positive roots (including $0$).
Let $\rho=\frac{1}{2}\sum_{\alpha\in R^+}\alpha$.
Let $P$ be the lattice of integral weights, $P^+$ be the set of
dominant integral weights. Let
$\delta(h)=\prod_{\alpha\in R^+}(e^{\langle \alpha,h\rangle
/2}-e^{-\langle \alpha,h\rangle /2})$, $h\in\frak h$,
be the Weyl denominator.
Let $\frak t\subset\frak h$ be the Lie algebra of $T$.
Let $e_{\alpha},f_{\alpha}$ be the root vectors corresponding
to the roots $\alpha$ and
$-\alpha$ such that $\langle e_{\alpha},f_{\alpha}\rangle =1$.
Let $Y_1,...,Y_r$ ($r=\text{rank} (K)$)
be generators of $\Cal Z(\frak g)$. Let $Y_1$ be the Casimir
element $Y_1=\sum_{i=1}^{\text{dim}\frak g}a_i^2$, where $a_i$ is
an orthonormal basis of $\frak g$ with respect to $\langle ,\rangle $.
Let $M_{\l}$ be the Verma module, and
$N_{\l}$ be any highest weight module over the Lie algebra
$\frak g$ with highest weight $\l$. Let $N_{\l}[\mu]$ denote
the subspace of weight $\mu$ in $N_{\l}$.
Let $(,)$ denote the contravariant form on $N_{\l}$. Let
$\Phi: N_{\l}\to N_{\l}\otimes V^*$ be an intertwining operator for
$\frak g$.

The calculation of the radial part gives the following result.

\proclaim{Proposition 1.4}

(i) (Harish-Chandra) The radial parts $R_V(Y_1),..., R_V(Y_r)$
are pairwise commutative differential operators in $r$ variables
whose symbols are the highest terms of the polynomials $\chi(Y_1),...,
\chi(Y_r)$.

(ii) (Harish-Chandra) The second order Laplace
operator $\Delta=R_V(Y_1)$ has the form
$$
(R_V(Y_1)\psi)(h)=\delta(h)^{-1} \biggl(\Delta_{\frak
h}-\sum_{\alpha\in R^+}\frac{e_{\alpha}f_{\alpha}}
{2\sinh^2(\langle \alpha,h\rangle /2)}-\langle \rho,\rho\rangle
\biggr)(\psi(h)\delta(h)), h\in{\frak t}\tag 1.3
$$
where $\Delta_{\frak h}$ is the  Laplace operator
on $\frak h$, and $\psi(h)=\tilde\psi(e^h),\tilde\psi\in F^V(K)$.

(iii) (\cite{E},\cite{ES}) Set
$$
\Psi(h)=\Tr|_{N_{\l}}(\Phi e^h)=
e^{\langle \l,h\rangle }\sum_{\alpha\in Q^+}\Tr|_{N_{\l}[\l-\alpha]}(\Phi)
e^{-\langle \alpha,h\rangle }.\tag 1.4
$$
This series absolutely converges in the region
$\{h\in{\frak h}:\text{Re}\langle \alpha,h\rangle >0, \alpha\in
R^+\}$,
 and its sum takes values in $V^*[0]$ and
is an analytic solution of the holonomic system of partial differential
equations
$$
R_V(Y_j)\Psi=\Lambda_j\Psi,\ \Lambda_j=\chi(Y_j)(\l),\ 1\le j\le r \tag 1.5
$$
in this region.
\endproclaim

In fact, for generic complex numbers $\Lambda_1,...,\Lambda_r$
solutions of (1.5) given by (1.4) span the space
of solutions of (1.5) with values in $V^*[0]$.
Namely, let $\l$ be such that the stabilizer of $\l+\rho$
 in the Weyl group is the identity, and the Verma modules
$M_{w(\l+\rho)-\rho}$ are irreducible for all $w\in W$.
Let $m=\text{dim}V^*[0]$. If $M_{\mu}$ is irreducible
then $\text{dim}(\text{ Hom}(M_{\mu},M_{\mu}\otimes V^*))=m$.
Let $\Phi_j(\mu)$, $j=1,...,m$, be a basis of
$\text{ Hom}(M_{\mu},M_{\mu}\otimes V^*)$, and
let $\Psi_{j\mu}(h)=\Tr|_{M_{\mu}}(\Phi_j(\mu) e^h)$.

\proclaim{Proposition 1.5} (see \cite{E})

(i) The functions $\Psi_{j\mu}$, $j=1,..,m$, $\mu=w(\l+\rho)-\rho$,
are linearly independent and span the space of $V^*[0]$-valued
solutions of (1.5).

(ii) If $M_{\mu}$ is an irreducible
module then for every $\Phi\in \text{Hom}(M_{\mu},M_{\mu}\otimes V^*)$
the function $\Psi(h)=\Tr|_{M_{\mu}}(\Phi e^h)$ can be characterized
as a solution of (1.5) of the form
$e^{\langle \mu,h\rangle }\Psi_0(h)$, where $\Psi_0$ is analytic
in the region $\{h: \text{Re}\langle \alpha,h\rangle >0,\alpha\in R^+\}$ and
tends to $(v_{\mu},\Phi v_{\mu})$ as $\langle \alpha,h\rangle \to +\infty$ for
all $\alpha\in R^+$.
\endproclaim

{\bf Remarks.} 1. In the theory of radial part one does
not essentially use the fact that $V^*$ is a finite-dimensional
representation of the group $K$. It is enough for $V^*$ to
to be any module over the Lie algebra $\frak g$
such that the space $V^*[0]$ is finite-dimensional.

2. All the above results easily generalize to the case
of an arbitrary compact Lie group. We are studying
the special case of a simply connected simple group
just for brevity.

The above results can be applied to the theory
of quantum integrable systems and to the theory of
symmetric polynomials, as follows.

\vskip .1in
{\bf 1.5. Integrability of the Sutherland operator.}

Let us fix $\ell\in\Bbb N$.
Consider the case $K=SU(n)$, $V^*=S^{(\ell-1)n}\C^n$.
In this case, the space $V^*[0]$ is one-dimensional, so
we assume that it is somehow identified with $\C$, by mapping $1\in\C$
to $u_0\in V^*[0]$. Then
the operators $R_V(Y)$ can be regarded as scalar differential operators.
It is convenient to realize $\frak h$ as a subspace
in $\C^n$ given by the equation $\sum_{i=1}^nx_i=0$, and set
$h=(x_1,...,x_n)$.
Since $e_{\alpha}f_{\alpha}|_{V^*[0]}=\ell(\ell-1)\text{Id}$, $\alpha\in R^+$,
and $\<\rho,\rho\>=\frac{n^3-n}{12}$,
the second order Laplacian (1.3) can be rewritten as follows:
$$
R_V(Y_1)=\delta^{-1}\circ
\biggl(\sum_{i=1}^n\frac{\d^2}{\d
x_i^2}-\ell(\ell-1)\sum_{i<j}\frac{1}
{2\text{sinh}^2(\frac{x_i-x_j}{2})}-\frac{n^3-n}{12}
\biggr)\circ\delta.
\tag 1.6
$$
Up to conjugation by $\delta$ and addition of a constant,
this is the Sutherland
operator -- the Hamiltonian of the quantum $n$-body
problem on the line with trigonometric potential.

\proclaim{Definition 1.4} Let $D$ be a differential operator
in $n$ variables. A differential
 operator commuting with $D$ is called a quantum
integral of $D$. One says that $D$ defines a completely integrable
quantum Hamiltonian system if there exists a set of $n$ algebraically
independent differential operators $D_1,...,D_n$ which
commute with $D$. The system $D_1,...,D_n$
is called a complete system of quantum integrals for $D$.
\endproclaim

Propositions 1.4,1.5 imply the following result:

\proclaim{Proposition 1.6}

(i) \cite{OP}
The Sutherland operator
$$
D_\ell=\sum_{i=1}^n\frac{\d^2}{\d
x_i^2}-\ell(\ell-1)\sum_{i<j}\frac{1}
{2\text{sinh}^2(\frac{x_i-x_j}{2})}
$$
 defines a completely integrable
quantum Hamiltonian system.

(ii) \cite{E} The quantum integrals of the Sutherland operator
are equal to $\delta\circ R_V(Y_j)\circ \delta^{-1}$, $j=1,..,r$.
The functions of the form $\delta\Psi$, where $\Psi$ is
given by (1.4), are common eigenfunctions
of these quantum integrals. For generic eigenvalues, functions
of this form  span the space of common eigenfunctions.
\endproclaim

\vskip .1in
{\bf 1.6. Jack's polynomials as spherical functions.}

Let us now introduce Jack's polynomials.
Consider the operator $\tilde D_\ell=\delta^{-\ell}\circ D_\ell\circ
\delta^{\ell}$.

\proclaim{Proposition 1.7} \cite{HO}
The operator $\tilde D_\ell$ maps the space
$\C[P]^W$ of Weyl group invariant Laurent polynomials
on the maximal torus into itself. Moreover, it is triangular with respect
to the basis of orbitsums $m_{\l}=\sum_{\nu\in W\l}e^{\langle \nu,h\rangle }$,
$\l\in P^+$, i.e.
$\tilde D_\ell
m_{\l}=\langle \l+\ell\rho,\l+\ell\rho\rangle m_{\l}+
\sum_{\nu<\l}c_{\l\nu}m_{\nu}$.
\endproclaim

This proposition implies that one can uniquely define a $W$-invariant
Laurent polynomial $J_{\l}^\ell=m_{\l}+\sum_{\nu<\l}s_{\l\nu}m_{\nu}$
by the condition $\tilde D_\ell J_{\l}^\ell=\langle \l+\ell\rho,\l+\ell
\rho\rangle J_{\l}^\ell$.
The polynomials $J_{\l}^\ell$ are called the Jack's polynomials.
They have the following interpretation in terms of spherical
functions.

Let $L_{\l}$ be the irreducible highest weight representation of $K$
corresponding to the dominant integral weight $\l$.
The following statement is checked directly.

\proclaim{Lemma 1.8}
 The space $\text{Hom}_K(L_{\mu},L_{\mu}\otimes V^*)$ is
one-dimensional if $\l=\mu-(\ell-1)\rho\in P^+$ and zero otherwise.
\endproclaim

 Let $\Phi_{\l}: L_{\l+(\ell-1)\rho}\to L_{\l+(\ell-1)\rho}\otimes V^*$ be
the intertwiner such that $(v_{\l},\Phi v_{\l})=u_0\in V^*[0]$, $\l\in P^+$.
Consider the conjugacy equivariant functions on $K$:
$\Psi_{\l}(x)=\Tr|_{L_{\l+(\ell-1)\rho}}(\Phi_{\l} x)$.
They belong to $\C[P]^W$.

\proclaim{Proposition 1.9}\cite{EK4} Let $x=e^h\in T$.
Then, under the identification $V^*[0]$ with $\C$:

(i) $\Psi_0(x)=\delta(h)^{\ell-1}$.

(ii) $\Psi_{\l}(x)$ is divisible by $\Psi_0(x)$ in the algebra
$\C[P]^W$ for any $\l\in P^+$.

(iii) $\frac{\Psi_{\l}(x)}{\Psi_0(x)}=J_{\l}^\ell(x)$.
\endproclaim

This together with Proposition 1.2 (ii) yields

\proclaim{Corollary 1.10} (\cite{HO,Hec,O1,O2,M})

(i) The Jack's polynomials
are orthogonal: $\int_{H}|\delta(x)|^{2\ell}J^\ell_{\l}(x)
J^\ell_{\nu}(x^{-1})dx=0$
if $\l\ne\nu$.

(ii) $J_{\l}^\ell$ are eigenfunctions of the quantum integrals
of the Sutherland operator with eigenvalues given by
the Harish-Chandra homomorphism.

(iii) The quantum integrals of $\tilde D_\ell$ are self-adjoint
with respect to the inner product $(f,g)\to
\int_H|\delta(x)|^{2\ell}f(x)g(x^{-1})dx$.
\endproclaim

\vskip .1in
\centerline{\bf 2. Spherical functions on affine Lie groups:
main results}
\vskip .1in

{\bf 2.1. Affine Lie algebras and groups.}

Let $G$ be a complex simply connected simple Lie group.
Let $K$ be a fixed maximal compact subgroup in $G$.
$K$ defines a real structure on $G$ and thus an antilinear
Cartan involution on the Lie algebra of $G$.

We denote by $LG$ the group of holomorphic maps from $\C^*$
to $G$ with pointwise multiplication. This group will be called
the loop group of $G$. We denote by $LK$ the subgroup of $LG$
consisting of all maps from $\C^*$ to $G$ for which the image
of the unit circle is in $K$.

Denote by $\hat G$ the universal central extension of $LG$.
This is a one-dimensional extension by $\C^*$; this
extension is a nontrivial holomorphic principal $\C^*$-bundle over
$LG$ whose first Chern class generates the group $H^2(LG,\Z)=\Z$
(see \cite{PS}). Also, the group $\C^*$ acts on $LG$ by $(q\circ
g)(z)=g(q^{-1}z)$; we denote the semidirect product
$\C^*\ltimes LG$ associated to this action by $\check G$.
The semidirect product $\C^*\ltimes\hat G$
is denoted by $\tilde G$. This group will play the role of the compact
Lie group considered in the previous Chapter.

Similarly, we define $\check K=S^1\ltimes LK$, $\hat K$ -- the
central extension of $LK$ by $S^1$, and $\tilde K=S^1\ltimes \hat K$,
as in \cite{PS}.

We will also use the Lie algebras of $G,LG,\hat G,\check G,\tilde G$,
which are denoted by $\g,L\g,\hat\g,\check \g,\tilde \g$.
The Lie algebra $L\g$ is called the loop algebra of $\g$.
By definition, it consists of all holomorphic
maps from $\C^*$ to $\g$, with the pointwise commutator;
the Lie algebra $\tilde \g$ is $L\g\oplus\C c\oplus \C d$, where
$$
[c,a(z)]=[c,d]=0,\ [d,a(z)]=za'(z),\ [a(z),b(z)]=[a,b](z)+\frac{1}{2\pi
i}\int_{|z|=1}\langle a'(z)b(z)\rangle dz\cdot c,
$$
where $[ab]$ denotes the pointwise commutator of $a$ and $b$, and
$\langle ,\rangle $ is the invariant form on $\g$ normalized in
 such a way that long
roots satisfy $\langle \alpha,\alpha\rangle =2$.
The Lie algebras $\hat \g$ and $\check \g$
are defined by $\hat \g=L\g\oplus\C c\subset \tilde \g$, $\check
\g=\tilde\g/\C c$.

Let $L\g_{pol}$ be the Lie subalgebra in $L\g$
consisting of the loops in $L\g$
expressed by $\g$-valued Laurent polynomials in $z\in\C^*$.
For $a\in\g$, $n\in \Z$, let $a[n]\in L\g_{pol}$ be defined by
 $a[n](z)=z^na$.
If $\{a_i\}$ is a basis of $\g$ then $\{a_i[n]\}$
is a basis of $L\g_{pol}$.
The corresponding extensions of $L\g_{pol}$ are
denoted by $\hat \g_{pol}$, $\check \g_{pol}$, $\tilde \g_{pol}$.

Let $\langle ,\rangle $ denote the
standard invariant form on $L\g$:
$$
\langle a(\cdot),b(\cdot)\rangle =\frac{1}{2\pi i}\int_{|z|=1} \<a(z),b(z)\>
\frac{dz}{z},
$$
This form extends to an invariant form on $\tilde \g$ by
setting $\<c,d\>=1$, $\<c,c\>=\<d,d\>=0$, $\<c,a(z)\>=\<d,a(z)\>=0$.

The Lie algebras $\hat \g_{pol}$, $\tilde \g_{pol}$ are
affine Kac-Moody Lie algebras, or shortly, affine Lie algebras; therefore,
we call $\hat G$, $\tilde G$ affine Lie groups.

\vskip .1in
{\bf 2.2. Spherical and equivariant functions.}

A direct application of the definition of a spherical
function given in Section 1.1
to the case of affine Lie groups does not give an interesting
result: there is no nontrivial examples. The reason for this is
that the set of conjugacy classes in $\tilde G$ as a whole
is geometrically unsatisfactory. However, parts of this set
have a very good geometric structure, which suggests
that there should be a good theory for spherical functions
defined on conjugacy invariant subsets of $\tilde G$.
We will see that such a theory indeed exists.

Let $q\in\C^*$. Denote by $\tilde G_q$ the subset in $\tilde G$
of elements of the form $(q, g)$, $g\in\hat G$. This set is invariant
under conjugation. We will study the case when $|q|<1$.

The spherical functions we consider are
holomorphic functions on $\tilde G_q$. Since $\tilde G_q$ is
infinite-dimensional, we have to define what it means.

\proclaim{Definition 2.1}

(i) Let $D$ be the unit disk in $\C$.
We say that a map $\phi: D\to LG$ is holomorphic
if the associated map $D\times \C^*\to G$ is a holomorphic function of
two variables.

(ii) We say that a function $f: LG\to \C$ is holomorphic
if for any holomorphic map $\phi:D\to LG$ the function $f(\phi): D\to\C$
is a holomorphic function of one variable.
\endproclaim

This allows one to define holomorphic functions on $\tilde G_q$
with values in a finite-dimensional complex vector space.

\proclaim{Definition 2.2}

(i) A holomorphic function $f:\tilde G_q\to \C$ is
called spherical if the space $V_f$ spanned by all the functions $f^g(y)=
f(g^{-1}yg)$, $g\in\hat G$, is finite-dimensional.

(ii) $f$ is called a central (class) function
if $V_f=\C$ as a $\hat G$-module.

(iii) Let $V$ be a finite-dimensional representation of $\hat G$.
A spherical function $f$ is of type $V$ if $V_f$ is isomorphic to $V$
as a $\hat G$-module.

(iv)
A holomorphic function $\Psi: \tilde G_q\to V^*$ is called
(conjugacy) equivariant if for any $g\in\hat G$, $x\in\tilde G_q$ one has
$g^{-1}\Psi(gxg^{-1})=\Psi(x)$.
\endproclaim

{\bf Remark. } The only differences between this definition and
Definitions 1.1-1.3 are that now the function $f$ is defined
on a subset of $\tilde G$ rather than on the whole group,
and one considers conjugations by elements of a subgroup $\hat G$ of $\tilde G$
rather than $\tilde G$ itself. The last change is necessary because
nontrivial finite-dimensional representations
of $\hat G$ do not extend to $\tilde G$.

 Let $z\in\C^*$ and $p_z:\hat G\to G$ be the
evaluation homomorphism -- the composition of
the projection $\hat G\to LG$ and the evaluation map $g(\cdot)\in LG\to
g(z)\in G$. Let $V$ be a finite-dimensional representation of $G$.
Define a $\hat G$-module $V(z)=p_z^*V$: $\pi_{V(z)}(g)=\pi_V(p_z(g))$.
This module is called an evaluation module.

\proclaim{Proposition 2.1} \cite{CP}

(i) Let $V_1$,...,$V_n$ be irreducible nontrivial $G$-modules.
Then the $\hat G$-module $V_1(z_1)\otimes\dots\otimes V_n(z_n)$
is simple iff $z_1,...,z_n\in\C^*$ are distinct.

(ii) Every nontrivial irreducible finite-dimensional representation
of $\hat G$ is isomorphic to $V_1(z_1)\otimes\dots\otimes V_n(z_n)$
for suitable $V_i\ne \C,z_i\ne 0$. $V_i$ and $z_i$ are determined uniquely
up to order.

\endproclaim

 From now on we assume that the representation $V$ is irreducible,
i.e. of the form described by Proposition 2.1.
The vector-function $\Psi_f$ corresponding to a spherical
function $f$ is defined as in Chapter 1, and Proposition 1.1
holds true (with $\hat G$ instead of the group $K$, and
$\tilde G_q$ instead of the symmetric space $K\times K/K_d$).

As we explained in Section 1.1, to study spherical functions of an
irreducible type is the same as to study equivariant functions.
So we will concentrate our attention on equivariant functions on
$\tilde G_q$.

\vskip .1in
{\bf 2.3. Equivariant functions of degree $k$ and their construction.}

Let $k\in\Z$. Let $Z_0$ be the connected component
of the identity in the center of $\hat G$. Let
$\zeta: Z_0\to \C^*$ be the identification
of $Z_0$ with $\C^*$.

We say that an equivariant function $\Psi$ is of degree
$k$ if $\Psi(zx)=\zeta(z)^k\Psi(x)$ for any $z\in Z_0$.
Let $F_k^V(\tilde G_q)$ denote the space of equivariant functions
$\Psi:\tilde G_q\to V^*$ of degree $k$ for any irreducible
finite-dimensional representation $V$ of
$\hat G$.

We say that $\Psi$ is a meromorphic equivariant function on $\tilde G_q$
of type $V$ and degree $k$ if $\Psi=u_1/u_2$, where
 $u_1: \tilde G_q\to V^*$ is a holomorphic equivariant function
of degree $l_1$, and
$u_2: \tilde G_q\to\C$ is a holomorphic central function of degree
$l_2$, so that $l_1-l_2=k$. It is clear that meromorphic equivariant
functions of a fixed degree form a vector space
(it is always infinite dimensional). We denote this space
by $MF^V_k(\tilde G_q)$. It can be shown that if an equivarinat function
of degree $k$ is defined and holomorphic almost everywhere,
and is locally (in a neighborhood of every point)
a ratio of two holomorphic functions, then it belongs to this space.

We would like to apply formula (1.2) to construct
equivariant functions on $\tilde G_q$. It turns out, however,
that $\tilde G$ does not have nontrivial finite-dimensional
representations. Thus, we will have to compute trace
in infinite-dimensional $\tilde G$-modules, namely, in highest weight
integrable modules. This requires some analytic background
which is introduced below.

Let  $L_{\l,k}$ denote the integrable $\tilde \g_{pol}$-module with highest
weight $\l$ and central charge $k$ (\cite{K1,PS}).
The action of the element $d$ in $L_{\l,k}$
is defined by the condition that it annihilates
 the highest weight vector. It was shown by H.Garland \cite{G} that
this module admits a
positive definite Hermitian form $(,)$,
contravariant with respect to the antilinear
Cartan involution on $\tilde \g_{pol}$ (this form should not be
confused with the {\it bilinear} contravariant form which we used in
Chapter 1). Let $v_{\l k\mu i}$ be an orthonormal basis
of the weight subspace $L_{\l,k}[\mu]$. We will consider
two completions of $L_{\l,k}$ defined in terms of this basis:
the $L^2$ completion
$$
L_{\l,k}^{ss}=\{\sum_{\mu,i}a_{\mu i}v_{\l k\mu i}:
\sum_{\mu,i}|a_{\mu i}|^2<\infty\}
$$
(ss=square summable), and the analytic completion
$$
L_{\l,k}^{an}=\{\sum_{\mu,i}a_{\mu i}v_{\l k\mu i}:
\text{ there exists }0<q<1\text{ such that }
a_{\mu i}=O(q^{-d(\mu)}), \mu\to\infty \},
$$
where $d(\mu)$ is the degree of the weight $\mu$ with respect to the
homogeneous grading (this degree takes nonpositive values).
Clearly, the $L^2$-completion is a Hilbert space,
and the analytic completion is a dense subspace in it. It is known
(see \cite{GW1}, Lemma 3.2)
that the action of $\tilde \g_{pol}$ on $L_{\l,k}$ extends
by continuity to an action of
$\tilde \g$ on $L_{\l,k}^{an}$ (but not on
$L_{\l,k}^{ss}$).

Introduce the semigroup $\tilde G_{<1}=\cup_{|q|<1}\tilde G_q$.
Let $d$ be the degree operator in $L_{\l,k}$, i.e.
$d|_{L_{\l,k}[\mu]}=d(\mu)$.

We will need the following results from the theory of loop groups:

\proclaim{Theorem 2.2}(\cite{GW1}, Theorem 6.1;
\cite{PS}) The action of the Lie algebra $\hat
\g$ in $L_{\l,k}^{an}$ uniquely integrates to an action of $\hat G$.
The restriction of this action to $\hat K$ is unitary with respect to
$(,)$.
\endproclaim

\proclaim{Lemma 2.3}
For any $q\in\C^*$ such that $|q|<1$
and any $g\in \hat G$ the operator $gq^{-d}:L_{\l,k}^{an}\to
L_{\l,k}^{an}$ extends to a trace class operator on $L_{\l,k}^{ss}$.
Thus, the representation of $\tilde\g$ in $L_{\l,k}$ exponentiates
to a representation of the semigroup $\tilde G_{<1}$ in $L_{\l,k}^{ss}$
by trace class operators.
\endproclaim

\demo{Proof} This lemma follows from \cite{GW1}, Theorem 6.1.
\enddemo

Let $V=V_1(z_1)\otimes V_2(z_2)\otimes\dots\otimes V_n(z_n)$, where
$V_i$ are irreducible finite-dimensional $\g$-modules, and $z_i$ are
distinct nonzero complex numbers, such that $|z_i|\ge |z_j|$ for $i<j$.
Let $Q(V)=|z_n/z_1|$.

Consider intertwining operators $\Phi: L_{\l,k}\to \hat
L_{\nu,k}\otimes V^*$, where $\hat L_{\nu,k}$ denotes the (formal) completion
of $L_{\nu,k}$ by degree. Denote the space of such intertwiners by
$I_{\l\nu k}(V^*)$. Set $\Phi(z)=(z^d\otimes 1)\Phi z^{-d}$.

\proclaim{Lemma 2.4}

(i) Let $q\in\C^*$, $|q|<Q(V)$. Then
for any intertwiner $\Phi\in I_{\l\nu k}(V^*)$ and any complex number $a$
such that $|q/z_n|<|a|<|1/z_1|$
the operator $\Phi(a) q^{-d}: L_{\l,k}\to \hat L_{\nu,k}\otimes V^*$
extends to a trace class operator $L_{\l,k}^{ss}\to L_{\nu,k}^{ss}\otimes
V^*$.

(ii) For any $g\in\hat G$ and $q$ as in (i) the operator
$\Phi(a)q^{-d}g: L_{\l,k}\to \hat L_{\nu,k}\otimes V^*$
extends to a trace class operator $L_{\l,k}^{ss}\to L_{\nu,k}^{ss}\otimes
V^*$.
\endproclaim

\demo{Proof} See Section 3.1.\enddemo

Now we can construct a large supply of equivariant functions on
$\tilde G_q$.

Let $\Phi: L_{\l,k}\to \hat L_{\l,k}\otimes V^*$ be an intertwiner.
Consider the operator $\Phi(a)q^{-d}g$, where $a$ is as in Lemma 2.4,
$g\in\hat G$, $|q|<Q(V)$. Since this operator is trace class, we can compute
its trace:
$$
\Psi(\tilde g)=\Tr|_{L_{\l,k}^{ss}}(\Phi(a)\tilde g)=
\Tr|_{L_{\l,k}^{ss}}(\Phi(a)q^{-d}g),\ \tilde
g=(q,g)\in \tilde G_q.\tag 2.1
$$
This function is obviously holomorphic, since the representation
$L_{\l,k}$ is a holomorphic representation. In particular,
if $V=\C$ and $\Phi=\text{Id}$, then (2.1) is just the
character of the module $L_{\l,k}$.

 From the cyclic property of the trace one gets:

\proclaim{Proposition 2.5} The function $\Psi$ defined by (2.1) is
conjugacy equivariant.
\endproclaim

{\bf Remark. } Insertion of $a$ in the operator $\Phi$ is a purely
technical point, since the trace (2.1), in which we are eventually
interested, does not depend on $a$.

Let $S(V,L_{\l,k})$ denote the space of functions on $\tilde G_q$
spanned by functions (2.1) for all possible choices of
the intertwiner $\Phi: L_{\l,k}\to \hat L_{\l,k}\otimes V^*$.
The spaces $S(V,L_{\l,k})$ are
finite-dimensional subspaces of $F_k^V(\tilde G_q)$, and they are
disjoint for distinct pairs $(\l,k)$. Also, the map assigning
the function (2.1) to an intertwiner $\Phi$ is
an isomorphism of vector spaces
$\text{Hom}_{\hat\g_{pol}}
(L_{\l,k},\hat L_{\l,k}\otimes V^*)\to S(V,L_{\l,k})$.

For $k>0$, let $P^+_k$ be the set of highest weights (with respect to $\g$) of
integrable $\hat \g$-modules at level $k$.
Define the space $H_k=\oplus_{\l\in P^+_k}L_{\l,k}\otimes L_{\l,k}^*$,
where $L_{\l,k}^*$ denotes the restricted dual space
to $L_{\l,k}$ (direct sum of the dual spaces to the weight subspaces
in $L_{\l,k}$).
This space is naturally a module with central charge $0$
with respect to the diagonal action of $\tilde \g_{pol}$.
In physics it arises as the space of states of the Wess-Zumino-Witten
conformal field theory at level $k$.

The following statement is an affine analogue of Proposition 1.2(ii).
It gives a classification of equivariant functions, which is one
of our main results.

\proclaim{Theorem 2.6}

(i) Any equivariant function of a negative degree is zero.

(ii) If $\Psi$ is a nonzero equivariant function of degree 0
then $V=\C$ and $\Psi$ is a constant.

(iii) If $k>0$ and $|q|$ is sufficiently small
then
$F_k^V(\tilde G_q)=\oplus_{\l\in P^+_k}S(V,L_{\l,k})=
\text{Hom}_{\hat \g_{pol}}(H_k,V^*)$.
\endproclaim

\demo{Proof} See Section 3.3.\enddemo

{\bf Remarks. } 1. Note that unlike the compact Lie group case,
the direct sum in Theorem 2.6 is finite for any $k$, so the
space $F_k^V(\tilde G_q)$ is finite-dimensional.
Thus, any equivariant function of degree $k$ can
be represented as a finite linear combination of functions of the form
(2.1).

2. The condition ``$|q|$ is sufficiently small'' can in fact be replaced
by a more precise condition ``$|q|<Q(V)$'', although our proof
does not show it. This refinement can be achieved by applying
a method suggested in Section 5 of \cite{FG}; see Section 3.3 for more
 details.

3. There exists a heuristic statement which is sometimes called
``Peter-Weyl theorem for affine Lie groups''. It says
that the space of $L^2$-functions on $\hat G$ of level $k$
is equal to $H_k=\oplus_{\l\in P_k^+}L_{\l,k}\otimes L_{\l,k}^*$.
This is not even a conjecture since the notion of an $L^2$-function
is not defined, but it is useful as a guiding principle in
mathematical understanding of conformal field theory. Theorem 2.6(iii)
(which is by itself a rigorous statement) can be regarded as
a consequence of this ``affine Peter-Weyl theorem'' in the same way as
Proposition 1.2(ii) is a consequence of the classical Peter-Weyl
theorem for compact Lie groups.

3. Theorem 2.6 has a nice interpretation in terms of
the WZW conformal field theory. The main structural
ingredient of a conformal field
theory is the modular functor, which was defined by G.Segal \cite{Se}.
This is a rule which assigns to every Riemann surface $\Sigma$ with
punctures and labels on them
a finite-dimensional vector space --
the space of conformal blocks on this surface.
If we take the surface $\Sigma$ to be the elliptic curve
$\C^*/q^{\Z}$, the punctures to be the projections of the points
$z_1,...,z_n$ to this curve, and the labels to be the highest weights
of the representations $V_1,...,V_n$ then the assignment
$(\Sigma,z_1,...,z_n,V_1^*,...,V_n^*)\to F_k^V(\tilde G_q)$
is the modular functor of the WZW theory at level $k$
(\cite{MS}). This completely describes the modular functor
for surfaces of genus 1.

\proclaim{Corollary 2.7} Characters of integrable modules
at level $k$ form a basis in the space of central functions
on $\tilde G_q$ of degree $k$ ($k>0$).
\endproclaim

\vskip .1in
{\bf 2.4. The second order Laplace operator on an
affine Lie group.}

 For $0<Q\le 1$, define
the set $\tilde G_{<Q}=\cup_{0<|q|<Q} \tilde G_q$.

\proclaim{Definition 2.3}

(i) A holomorphic function $\Psi: \tilde
G_{<Q}\to V^*$ is called an equivariant function of degree
$k$ if it is such after restriction to $\tilde G_q$ for any $q$ such
that $0<|q|<Q$. The space of equivariant
$V^*$-valued functions of degree $k$ on $\tilde G_{<Q}$
is denoted by $F_k^V(\tilde G_{<Q})$.

(ii) A meromorphic equivariant function on $\tilde G_{<Q}$ of type $V$
and degree $k$ is a ratio $u_1/u_2$, such that
$u_1\in F_{l_1}^V(\tilde G_{<Q})$, $u_2\in F_{l_2}^\C(\tilde G_{<Q})$,
and $l_1-l_2=k$. The space of such functions is denoted by
$MF_k^V(\tilde G_{<Q})$.
\endproclaim

Now we would like to define the Laplace operator on $\tilde G$.
For this purpose we recall that any element of $\tilde \g$
defines a left-invariant holomorphic vector field on $\tilde G$.
Therefore, a quadratic expression over $\tilde\g$ would
define a left-invariant second-order differential operator on $\tilde G$.
If we would like to obtain a two-sided invariant
operator, we must take an expression which commutes with $\tilde \g$.
Such an expression is known -- it is the Sugawara expression
for the Casimir element of $\tilde \g$. This expression belongs to a
completion of the universal enveloping algebra $U(\tilde \g)$.

\proclaim{Definition 2.4} The Laplace operator for $\tilde G$
is the differential operator given by the Sugawara expression:
$$
\Delta_{\tilde G}=D_C,\ C=2(c+h^{\vee})d+\sum_{a\in B}\sum_{n\in\Z}:a[n]a[-n]:,
$$
where $h^{\vee}$ is the dual Coxeter number of $G$,
$B$ is an orthonormal basis of $\g$ with respect to
$\langle ,\rangle $,  $:a[n]a[-n]:$
is the normal ordered product which equals $a[n]a[-n]$ if
$n<0$ and $a[-n]a[n]$ otherwise.
\endproclaim

Of course, Laplace operator is an infinite expression, and therefore
it cannot be applied to an arbitrary meromorphic function. Define
truncated Laplacians
$$
\Delta_{\tilde G}^{(N)}=D_{C_N},
C_N=2(c+h^{\vee})d+\sum_{a\in B}\sum_{n=-N}^N:a[n]a[-n]:\tag 2.2
$$
Let us say that a meromorphic function $f$ on $\tilde G_{<Q}$
with values in a finite-dimensional vector space
is admissible if there exists a pointwise limit $\Delta_{\tilde G}f=
\lim_{N\to\infty}\Delta_{\tilde G}^{(N)}f$ at every regular point of $f$.

\proclaim{Theorem 2.8} Let $0<Q<1$ be sufficiently small. Then:

(i) Meromorphic equivariant functions of degree $k$ on $\tilde G_{<Q}$
are admissible.

(ii) The spaces $F_k^V(\tilde G_{<Q})$, $MF_k^V(\tilde G_{<Q})$
are invariant under the Laplace
operator.

(iii) The Laplace operator is conjugacy invariant, i.e.
commutes with the action of $\hat G$ by conjugation
on $MF_k^V(\tilde G_{<Q})$.

(iv) Traces of intertwiners given by (2.1) are eigenfunctions
of the Laplace operator. That is, if $\Psi$ is given by
(2.1) then $\Delta_{\tilde G}\Psi=\langle \l,\l+2\rho\rangle \Psi$.
\endproclaim

\demo{Proof} See Section 3.4.\enddemo

\vskip .1in
{\bf 2.5. The radial part of the Laplace operator.}

Let us now consider the affine analogue of the theory of radial part.

Our notations for the root system for $G$, the Weyl group, etc.
are the same as we introduced in Chapter 1 for $K$.
Let $\xi_j,1\le j\le r$, be an orthonormal
 basis of the Cartan subalgebra $\frak h\subset \g$.

Let $\hat R$ be the set of roots of the affine Lie algebra
$\hat\g$, and let $\hat R^{\pm}$ be the sets of positive and
negative roots, respectively. Let $\tilde{\frak h}=\frak
h\oplus\C c\oplus\C d$ be the Cartan subalgebra of
$\tilde\g$. Let $\hat P\subset \tilde\frak h^*$ be the integral
weight lattice of $\hat R$, i.e the set of weights $\l+kc^*+ld^*$,
$k,l\in\Z$, $\l\in P$, where $c^*(h)=\<c,h\>$, $d^*(h)=\<d,h\>$,
$h\in\tilde\frak h$.
Let $\hat P_k\subset \hat P$ be
the set of weights of level $k$, $\hat
P_k^+\subset \hat P_k$ be the set of dominant integral weights.
Let $E_{\hat\alpha}^i$,
$F_{\hat\alpha}^i$ be bases of the root subspaces
$\hat\g_{\ha}$, $\hat\g_{-\ha}$, respectively, $\ha\in\hat R^+$, such that
$\langle E_{\hat\alpha}^i,
F_{\hat\alpha}^j\rangle =\delta_{ij}$ (recall that the root subspaces
are not always one-dimensional for imaginary roots).
Then the Casimir element for $\tilde \g$ can be written as follows:
$$
C=2(c+h^{\vee})d+\sum_{j=1}^r \xi_j^2+2\rho+\sum_{\ha\in\hat R^+}
F_{\ha}E_{\ha}.
$$

As in the finite-dimensional case (see Chapter 1), any conjugacy
equivariant function on $\tilde G_{<Q}$
is completely determined by its values on the subset
$\{(q,g)| g\in H\times\text{ center}\}$, where $H$ is the Cartan
subgroup in $G$. This is a consequence of the fact that
the set of elements of $\tilde G_{<Q}$ which can be mapped into this
set by conjugations by elements of $\hat G$ is a (Zariski) dense
 set in $\tilde G_{<Q}$ (see Section 3.2). Further, if an equivariant
function is of degree $k$, it is uniquely defined
by its values on $\{(q,g)| g\in H\}$. Thus, any
differential operator $D_Y$ on $V^*$-valued
conjugacy equivariant functions
of degree $k$ can be written in terms of the derivatives $\d/\d q$,
$\d/\d h$, $h\in\frak h$. We will call this operator acting on
functions of $q,h$ the radial
part of $D_Y$ and denote it by $R_V(Y)$ (here $Y$ is from a completion
of $U(\tilde \g)$).

The radial part of Laplace operator is given by a formula that
is completely analogous to the finite-dimensional formula (1.3).

Let $\hat h=h+td$, $t\in\C$, $q=e^{-t}$. Let $\hat\delta$
be the affine Weyl denominator:
$\hat \delta(\hat h)=e^{\langle \hat\rho,\hat
h\rangle }\prod_{\ha\in\hat R}(1-e^{-\langle \ha,\hat h\rangle })$, where
$\hat \rho=\rho+h^{\vee}c^*$.

\proclaim{Proposition 2.9}

 The operator $R_V(C)$ has the form
$$
(R_V(C)\psi)(\hat h)=\hat\delta(\hat h)^{-1}
\biggl(2(k+h^{\vee})\frac{\d}{\d t}+\Delta_{\frak
h}-\sum_{\hat\alpha\in\hat R^+}\frac{E_{\hat\alpha}F_{\hat\alpha}}
{2\sinh^2(\langle
\hat\alpha,\hat h\rangle /2)}-\langle \rho,\rho\rangle \biggr)(\psi(\hat
h)\hat\delta(\hat
h)),\tag 2.3
$$
where $\Delta_{\frak h}$ is the  Laplace operator
on $\frak h$.
\endproclaim

\demo{Proof} This statement follows from \cite{EK1}, Theorem 4.1; see also
\cite{Be},\cite{FG}.

In particular, if
 $V=V_1(z_1)\otimes\dots\otimes V_n(z_n)$, then
(2.3) can be rewritten in a more explicit form. Namely, for
every $b\in\g$ let $b_i$ denote the operator in $V^*$
obtained by the action of $b$ in the $V_i^*$-component of the tensor
product. Also, consider the functions
$$
\gather
\varphi(x,z,t)=\sum_{m\in\Z}\frac{z^m}{\sinh^2((x-mt)/2)},\\
\varphi_0(z,t)=\sum_{m\in\Z\setminus {0}}\frac{z^m}{\sinh^2(mt/2)}=
\lim_{x\to 0}\biggl(\varphi(x,z,t)-\frac{1}{\sinh^2(x/2)}\biggr).\tag 2.4
\endgather
$$
These functions can be expressed in terms of standard elliptic functions
as follows:
$$
\gather
\varphi(x,z,t)=-4\frac{\d}{\d x}\frac{
\Theta(x+\zeta)\Theta'(0)}{\Theta(x)\Theta(\zeta)},\\
\varphi_0(z,t)=\frac{2\Theta''(\zeta)}{\Theta(\zeta)}-\frac{2\Theta'''(0)}
{3\Theta'(0)}+\frac{1}{3},\endgather
$$
where $z=e^\zeta$, and

$$\Theta(x)=2q^{1/8} \sinh(x/2) \prod_{n\ge 1}
	(1-e^xq^n)(1-e^{-x} q^n)(1-q^n)
$$
is the standard theta-function.

\proclaim{Proposition 2.10}

Let $\{e_{\alpha},f_{\alpha},\xi_m\}$ be the root basis of $\g$.
If $V=V_1(z_1)\otimes\dots\otimes V_n(z_n)$
then the operator $R_V(C)$ can be written in the form
$$
\gather
(R_V(C)\psi)(\hat h)=\\
\hat\delta(\hat h)^{-1}
\biggl(2(k+h^{\vee})\frac{\d}{\d t}+\Delta_{\frak
h}-\frac{1}{2}\sum_{i,j=1}^n\sum_{\alpha\in R^+}\varphi
(\langle \alpha,h\rangle ,
\frac{z_i}{z_j},t)(e_{\alpha})_i(f_{\a})_j\\
-\frac{1}{4}\sum_{i,j=1}^n\sum_{m=1}^r\varphi_0(\frac{z_i}{z_j},t)
(\xi_m)_i(\xi_m)_j-\<\rho,\rho\rangle \biggr)
(\psi(\hat
h)\hat\delta(\hat
h)),\tag 2.5\endgather
$$
where $\psi(\hat h)=\tilde\psi(e^{\hat h}),
\tilde\psi\in MF_k^V(\tilde G_{<Q})$.
\endproclaim

\demo{Proof} This proposition is proved in
\cite{EK1}. It also follows from Proposition 2.9.\enddemo

{\bf Remarks. } 1. A commutative set of $r+2$ operators for which functions of
the form (2.1) are eigenfunctions can be obtained at any value of $k$.
They can be constructed as radial parts of Laplace operators
defined by central elements lying in a completion of
$U(\tilde \g)$. Two such elements are obvious --
the element $c\in\tilde\g$, and the Casimir
element $C$. However, the rest of these central elements
 (which were constructed in \cite {K2}) are quite complicated.
Their radial parts are not differential operators but rather
formal series in $q$ whose coefficients are differential operators.

2. The parabolic differential operator on the right hand side of (2.5)
coincides (up to conjugation by the affine Weyl denominator)
with the operator representing the Knizhnik-Zamolodchikov connection on
conformal blocks on an elliptic curve corresponding to the deformation
of the complex structure on this curve \cite{Be}.

\vskip .1in
{\bf 2.6. Affine Jack polynomials.}

Now we can define the affine analogue of Jack's polynomials, following
the paper \cite{EK3}, by analogy with Section 1.6.
 Consider the case when $\g=\sln,
V^*=S^{(\ell-1)n}\C^n(z)$. In this case, the trace $\Psi$ defined by
(2.1), when restricted to the Cartan subgroup,
 takes values in $V^*[0]$, which is one-dimensional and therefore
can be identified with $\C$. Also, it does not depend on $z$. Since,
as we have mentioned, $(e_\a f_\a)_{|V^*[0]}=\ell(\ell-1)\text{Id}$,
we can rewrite the
radial part of the Laplacian $R_V(C)$ as an operator acting on scalar
functions by the formula $R_V(C)=\dhat^{-1}\circ (\hat D_\ell -\langle \rho,
\rho\rangle )\circ \dhat$, where

$$\hat D_\ell = 2(k+h^{\vee})\frac{\d}{\d t}+\Delta_{\frak
h}-\frac{\ell(\ell-1)}{2}\sum_{\alpha\in R^+}\varphi
(\langle \alpha,h\rangle ,1,t).\tag
2.6$$

Define the operator $\tilde D_\ell= \dhat^{-\ell}\circ \hat D_\ell\circ
\dhat^{\ell}$. Let $\hat W$ be the affine Weyl group for $\g$, and let
$m_{\hat \l}$ be the orbitsums for $\hat W$:
$m_{\hat\l}=\sum_{\hat\nu\in \hat W\hat \l}e^{\langle
\hat \nu,\hat h\rangle }$.
By definition, a $W$-invariant
theta function of level $k$ is a
(possibly infinite) linear combination of orbitsums $m_{\hat \l}$ for
$\hat\l\in \hat P^+_k$, of the form
$\sum_{\hat\mu\le\hat\l}c_{\hat\l\hat\mu}
m_{\hat\mu}(\hat h)$,
which is convergent in the region $\text{Re} t >0$ ($\hat
h=h+td$, so $t=d^*(h)$), see \cite{Lo}. The space of such functions is
denoted by $\CK$. It is known (\cite{Lo}, \cite{BS})
that $\CK$ coincides with the space of
holomorphic functions $f(q,\xi ,u)$ of the variables $(q,\xi ,u)$
($|q|<1, \xi\in H, u\in\C^*$) such that $f$ is representable in the
form $f(q,\xi ,u)=f_0(q,\xi )u^k$, and invariant with respect to the action
of $\hat W$ given by equation (3.3) below.

Then we have an affine analogue of Proposition 1.7.

\proclaim{Proposition 2.11} (\cite{EK3})

The operator $\tilde D_\ell$ maps the space of $\hat W$-invariant theta
functions of any level $k\ge 0$
 into itself. Moreover, its action  is triangular with
respect to the basis of orbitsums: $\tilde D_\ell m_\lhat= \langle \lhat
+\ell\rhat, \lhat+\ell\rhat\rangle m_\lhat+\sum_{\hat\nu<\hat\l}
c_{\hat\l\hat\nu}m_{\hat\nu}$.
\endproclaim

Now, let us define affine Jack's polynomials $\hat J_\lhat^\ell$ as level
$k$ $W$-invariant theta functions such that

1. $\hat J_\lhat^\ell (\hat h) =e^{\langle \lhat, \hat h\rangle }+
\sum_{\hat\mu<\lhat}
s_{\lhat\hat\mu} e^{\langle \hat\mu, \hat h\rangle } $

2. $\tilde D_\ell\hat J_\lhat^\ell=\langle \lhat+\ell\rhat,
\lhat+\ell\rhat\rangle \hat J_\lhat^\ell$.

Then we have the following way to construct these polynomials. Let us
fix $\ell\in \Z_+$ and consider intertwiners
$\Phi_\lhat (z):L_{\lhat+(\ell-1)\rhat} \to
\hat L_{\lhat+(\ell-1)\rhat}\otimes V^*$, where $V^*$ is as above. Such an
 intertwiner exists iff
$\lhat=\lambda+kc^*+ld^*\in \hat P^+$, or $k\in \Z_+,l\in\Z,
\lambda\in P^+_k$.
Let us denote
$\Psi_\lhat(\tilde g)=\Tr|_{L_{\lhat+(\ell-1)\rhat}}
 (\Phi_\lhat(a)\tilde g)$. This is an equivariant function on
$\tilde G_{<Q}$ of level $k+(\ell-1)h\v$.
Consider its restriction to the elements of the Cartan subgroup in
$\tilde G$. This restriction belongs to the algebra $\CK$, and
one has the following theorem, which is an
affine analogue of Proposition 1.9:

\proclaim{Theorem 2.12}(\cite{EK3})
 Let $\tilde g=e^{\hat h}$, $\hat h=h+td$, $h\in\frak
h$. Then

(i) $\Psi_0(\tilde g)=\dhat^{\ell-1}(\hat h)$.

(ii) $\Psi_\lhat$ is divisible by $\Psi_0$ in the algebra
$\oplus_{k\ge 0}\CK$ for every $\lhat\in \hat P^+$
(i.e. the ratio is a theta function).

(iii) $\frac{\Psi_\lhat(\tilde g)}{\Psi_0(\tilde g)}= \hat  J_\lhat(\hat h)$.

\endproclaim

\vskip .1in
{\bf 2.7. Higher Laplace operators at the critical level
and their radial parts.}

We have described an affine analogue of the theory of radial part.
The difference with the finite-dimensional case is that we obtained
only one differential operator whereas in the finite-dimensional
case there is a
set of $\text{rank}\g$ pairwise commuting differential operators. It turns out
that a commuting set of differential operators can also be
obtained in the affine
case if we restrict the above construction to the critical level
$k=-h^{\vee}$.

Since the critical level $k=-h^{\vee}$ is negative, there is no
nonzero holomorphic equivariant functions of degree $k$ on $\tilde G_q$.
However, there are meromorphic equivariant functions.
We will construct a system
of $r=\text{rank} \g$ commuting differential operators
acting on these functions. This construction uses the existence
of $r$ algebraically independent central elements
of degree zero in a completion of $U(\hat \g)/(c=-h^{\vee})$, which
was proved by B.Feigin and E.Frenkel \cite{FF}. We explain this result below.

Let $U_k(\hat\g)=U(\hat \g)/(c=k)$.
Let $a\in\g$. Define the ``quantum field'' $a(z)=\sum_{m\in\Z}a[m]z^{-m-1}$
(it is just a formal series). We call such a field a basic field.
We can construct more complicated fields from basic fields
using addition, multiplication by numbers, differentiation by $z$,
and normal ordered product. Laurent components of such fields will
already be infinite expressions over $U_k(\hat\g)$, but they will give rise to
well defined operators in highest weight modules over $\hat\g$ with
central charge $k$. Denote the space of all fields obtained
in this way by $N_{0,k}$. As a vector space,
it coincides with the ``Weyl module'' over $\hat\g$
whith highest weight $0$ and central charge $k$.
If $k\ne -h^{\vee}$, then $N_{0,k}$ is a vertex operator algebra
(cf. \cite{FZ}). For $k=-h^{\vee}$, all axioms of a vertex operator
algebra hold excluding the axioms involving the Virasoro algebra
(since the Virasoro algebra action in $N_{0,k}$ has a pole at $k=-h^{\vee}$).
An element $A(z)\in N_{0,k}$ is called a field (a vertex operator)
of conformal dimension $\Delta$ if $[d,A(z)]=(-z\d-\Delta)A(z)$,
$\d=\frac{d}{dz}$. For instance, basic fields are of conformal
dimension $1$.

Fourier components of vertex operators from $N_{0,k}$ span a
space of infinite expressions which turns out to be closed under
commutation, i.e. a Lie algebra. Denote this Lie algebra
by $U_k(\hat\g)_{loc}$ (``quantized local functionals''), as in \cite{FF}.

Let $d_1,...,d_r$ be the exponents of $\g$. Let $\{a_j\}$ be an
orthonormal basis of $\g$.

\proclaim{Proposition 2.13}(\cite{FF}) There exist fields
$Y_1(z)=\sum_{j=1}^{\text{dim} \g}:a_j^2(z):$,
$Y_2(z),...,Y_r(z)$ $\in N_{0,-h^{\vee}}$ of conformal dimensions
$d_1,...,d_r$ such that their Fourier components commute with $\hat
\g$, are linearly independent, and span the center
of the Lie algebra $U_{-h^\vee}(\hat\g)_{loc}$.
\endproclaim

{\bf Remark. } Weaker versions of this result were obtained earlier
in \cite{Ma},\cite{GW2},\cite{Ha}.

Set $\hat Y_i=\frac{1}{2\pi i}
\oint Y_i z^{d_i-1}dz$. Then $\hat Y_i$ have degree
zero, i.e. commute with $d$. Define the corresponding differential
operators $D_{Y_i}=\Delta_i$ of orders $d_i$ on $\hat G$. We have:
$\Delta_1=\Delta_{\tilde G}$ (see Definition 2.4), and
$[\Delta_i,\Delta_j]=0$ on functions of degree $-h^{\vee}$.
Notice that $\hat Y_1=C$ -- the Sugawara element.

As we have seen, the operators $\Delta_i$ do not act on all
holomorphic or meromorphic
functions since they are defined by infinite expressions.
However, one can define the notion of an admissible function
for $\Delta_i$ similarly to the case of $\Delta_1$: one defines
truncated Laplacians $(\Delta_i)^{(N)}$ and calls a function
$f$ admissible if the sequence $(\Delta_i)^{(N)}f$ is convergent at every
regular point of $f$.

\proclaim{Theorem 2.14}

(i) Equivariant meromorphic functions are admissible for $\Delta_i$.

(ii) The space $MF_{-h^{\vee}}^V(\tilde G_q)$ is invariant under
$\Delta_i$.

(iii) $\Delta_i$ are conjugacy invariant.
\endproclaim

\demo{Proof} See Section 3.4.\enddemo

Since any equivariant function is uniquely determined by its
restriction to the Cartan subgroup, one can rewrite the
operators $\Delta_1,...,\Delta_r$ in the Cartan coordinates.
This will produce their radial parts, $R_V(\hat Y_1),...,R_V(
\hat Y_r)$. Using the structure of the elements $\hat Y_i$
(\cite{FF}) and Proposition 2.10, one obtains the following result.

\proclaim{Theorem 2.15}

(i) The radial parts
$R_V(\hat Y_1),..., R_V(\hat Y_r)$
are pairwise commutative differential operators in $r$ variables
whose symbols are the highest terms of the polynomials $\chi(\hat Y_1),...,
\chi(\hat Y_r)$, where $\chi: \hat U(\tilde\g)\to U(\frak h)$ is the
Harish-Chandra homomorphism ($\hat U(\tilde\g)$ denotes the completion
of $\tilde\g$).

(ii) The operator $R_V(\hat Y_1)$ for $V=V_1(z_1)\otimes\dots\otimes
V_n(z_n)$ has the form
$$
\gather
(R_V(C)\psi)(\hat h)=\\
\hat\delta(\hat h)^{-1}
\biggl(\Delta_{\frak
h}-\frac{1}{2}\sum_{i,j=1}^n\sum_{\alpha\in R^+}\varphi(\langle
\alpha,h\rangle ,
\frac{z_i}{z_j},t)(e_{\alpha})_i(f_{\a})_j\\
-\frac{1}{4}\sum_{i,j=1}^n\sum_{m=1}^r\varphi_0(\frac{z_i}{z_j},t)
(\xi_m)_i(\xi_m)_j-\langle \rho,\rho\rangle \biggr)
(\psi(\hat
h)\hat\delta(\hat
h)).\tag 2.7\endgather
$$
where $\Delta_{\frak h}$ is the  Laplace operator
on $\frak h$.
\endproclaim

\vskip .1in
{\bf 2.8. Integrability of the elliptic Calogero-Sutherland-Moser Hamiltonian.}

It is especially interesting to specialize Theorem 2.15 to the case
when $\g=\sln$, $V^*=S^{(\ell-1)n}\C^n(z)$. In this case, equivariant
functions restricted on the Cartan subgroup take values
in $V^*[0]=\C$, so the operators $R_V(\hat Y_i)$ can be regarded as
scalar differential operators. In particular,
 we can rewrite the
radial part of the Laplacian $R_V(\hat Y_1)$ as an operator acting in scalar
functions by the formula $R_V(\hat Y_1)=\dhat^{-1}\circ (\hat D_\ell -
\langle \rho,
\rho\rangle )\circ \dhat$, where

$$\hat D_\ell = \Delta_{\frak
h}-\frac{\ell(\ell-1)}{2}\sum_{\alpha\in R^+}\varphi
(\langle \alpha,h\rangle ,1,t).\tag
2.8
$$

On the other hand,
it is easy to check that
$\varphi(x,1,t)=-\frac{1}{\pi^2}\wp(\frac{x}{2\pi i},\tau)+c(\tau)$,
where $\tau=it/2\pi$,
$\wp$ is the Weierstrass elliptic function and
$c(\tau)=8\sum_{m>0}\frac{q^m}{(1-q^m)^2}-\frac{1}{3}$.
Therefore, up to renormalization,
operator (2.8) coincides with the Calogero-Sutherland operator
with elliptic potential \cite{OP}. Let us realize $\frak h$ as the
space $\{(x_1,...,x_n)\in\C^n:\sum_{i=1}^nx_i=0\}$, and set
$h=2\pi i(x_1,...,x_n)$. Then, from
Theorem 2.15 we obtain the following theorem.

\proclaim{Theorem 2.16} (i) (\cite{OP})
The elliptic Calogero-Sutherland operator
 $$
D_\ell=\sum_{i=1}^n\frac{\d^2}{\d
x_i^2}-\ell(\ell-1)\sum_{i\ne j}\wp({x_i-x_j},\tau)\tag 2.9
$$
defines a completely integrable
quantum Hamiltonian system.

(ii) (\cite{E}) The quantum integrals of $D_{\ell}$
are equal to $\hat\delta\circ R_V(\hat Y_j)\circ \hat\delta^{-1}$, $j=1,..,r$.
\endproclaim
\vskip .in

 Since the operators $R_V(\hat Y_i)$ are
pairwise commutative, one can consider the holonomic system of
differential equations $R_V(\hat Y_j)\Psi=\Lambda_j\Psi$, which is an
elliptic analogue of system (1.5). For any set of eigenvalues
$\Lambda_1,...,\Lambda_r$  there exist $|W|$ linearly independent
solutions of this system, which are expressed by interesting
special functions. For example, for $\g=\frak{sl}_2$ one has
only one second order differential
equation in one variable. This equation coincides with the classical
Lam\'e equation, and its solutions are the classical Lame
functions \cite{WW}. Thus, Lam\'e functions can
be interpreted as spherical functions on $\tilde{SL}_2$
of degree $-2$ which are eigenfunctions for the second order Laplace
operator.

{\bf Remark. }
  It is a natural question if one can give a representation theoretic
construction for solutions of the system $R_V(\hat Y_j)\Psi=\Lambda_j\Psi$,
similar to formula (1.4). One could expect that such a formula
would involve the trace of the operator $\Phi e^{\hat h}$ (where $\Phi$
is an intertwiner) in a module $N$ over $\hat\g$ at the critical level
$k=-h^{\vee}$. Unfortunately, this runs into the following difficulty:
nontrivial intertwining operators $\Phi: N\to \hat N\otimes V^*$
at the critical level do not exist. However, one can consider
the trace $\Tr|_N(\Phi e^{\hat h})$ for modules $N$ of level $k\ne
-h^{\vee}$, and study its asymptotics as $k\to -h^{\vee}$.
This was done in \cite{EK1} for $\g={\frak sl}_2$,
and it was shown that the leading
term of this asymptotics gives Lam\'e functions.
This result can be generalized to the case $\g={\frak sl}_n$,
which allows to obtain a representation-theoretic interpretation
of eigenfunctions of the elliptic Calogero-Sutherland-Moser system.
This will be described in a future paper.

\centerline{\bf 3. Proofs}

\vskip .1in
{\bf 3.1. Proof of Lemma 2.4.}
(i) Let $p$ be a complex number
such that $|p|>|q|$, $|p/z_n|<|a|<|1/z_1|$. It is enough to show that
the operator $\Phi(a)p^{-d}$ is bounded.
Pick a vector $w\in V$ and let $A: L_{\l,k}\to L_{\nu,k}$
be defined by the formula $Ax=(w,\Phi(a)p^{-d}x)$.
We must show that $A$ is bounded.
For this, it is enough to show that $\Tr(A^*A)$ is finite.
We do it by computing this trace explicitly.

Let $\theta: G\to G$ be the compact involution -- the antiholomorphic
involution such that $G^\theta =K$. This involution defines an antiholomorphic
involution $\hat\theta: LG\to LG$: $\hat\theta(g)(z)=\theta(g(1/\bar z))$,
such that $LG^{\hat\theta}=LK$. This involution extends to $\tilde G$
in an obvious way. The representation
$\pi:\tilde G\to \text{End}L_{\l,k}^{an}$
has the property: $\pi(\hat\theta(g))^{-1}=\pi(g)^*$, where
star denotes Hermitian conjugation with respect to the sesquilinear
contravariant form.

Let $V^\theta$ denote the representation
$V$ twisted by the involution $\theta$.
The main idea is to observe that the operator
$\Phi^*: L_{\nu,k}\otimes V^{\theta}\to L_{\l,k}$, Hermitian conjugate
to $\Phi$, is also an intertwiner, because the positive definite Hermitian form
on $L_{\l,k}$ is contravariant. Also, we have $\Phi(z)^*=\Phi^*(\bar z^{-1})$.
Therefore, $A^*A$ is a component of the operator
$X=\bar p^{-d}\Phi^*(\bar a^{-1})\Phi(a)p^{-d}: L_{\l,k}\to
L_{\l,k}\otimes V^*\otimes V^{\theta *}$ (here we regard
$\Phi^*$ as an operator $L_{\nu,k}\to L_{\l,k}\otimes V^{\theta *}$).

Our task is to show that the trace $\Tr(X)$ is finite.
Let $P_m$ be the orthogonal projector in $L_{\l,k}$ to the subspace
of vectors of degree $-m$. Let $X_m=P_mXP_m$, and let $T_m=\Tr(X_m)$
(it is a finite positive number). Then
$\Tr(X)=\sum_{m\ge 0}T_mp^m\bar
p^m=\sum_{m\ge 0}T_m|p|^{2m}$. We must show that this power series is
convergent.

We have
$$
\Tr(X)=\Tr(\Phi^*(\bar a^{-1})\Phi(a)|p|^{-2d})
.\tag 3.1
$$
But this is a correlation function for the WZW model on the torus,
for which the finiteness (i.e. convergence of the series)
follows from the inequalities $|z_n/p|^2>|a|^{-2}>|z_1|^2$
(=the definition of $a$) and from

\proclaim{Lemma 3.1}

Let $V^1,V^2$ be irreducible finite-dimensional representations
of $\hat G$, $V^1=V_1^1(z_1)\otimes\dots\otimes V_n^1(z_n)$,
$V^2=V_1^2(w_1)\otimes\dots\otimes V_m^2(w_m)$, so that $|z_i|\ge |z_j|$,
$|w_i|\ge |w_j|$, $i<j$. Let
$\Phi_1:L_{\l,k}\to L_{\nu,k}\otimes V^1$,
$\Phi_2:L_{\nu,k}\to L_{\l,k}\otimes V^2$ be intertwining operators.
Then the formal series
$$
F(x_1,x_2,q)=\Tr(\Phi_1(x_1)\Phi_2(x_2)q^{-d})=\sum_{m\ge 0}\Tr(P_m
\Phi_1(x_1)\Phi_2(x_2)P_m)q^m\in V^1\otimes V^2
\tag 3.2
$$
absolutely converges in the region $|w_n/qz_1|>|x_1/x_2|>|w_1/z_n|$.
\endproclaim

This Lemma follows from the fact that trace (3.2)
satisfies the Knizhnik-Zamolodchikov differential equation
(see \cite{EK1},\cite{Be},\cite{FG}) whose coefficients
are elliptic functions of $x_1/x_2$, and
whose Laurent expansions converge in the
region specified in the Lemma 3.1. (see \cite{EK1} for details)
This implies Lemma 2.4(i).

Statement (ii) of Lemma 2.4. follows from the facts
that 1) the operator $\Phi(a)q^{-d}g$ can be written as
$\Phi(a)p^{-d}g_1(q/p)^{-d}$, where $g_1=(q/p)^{-d}g(q/p)^d$,
2) the operators $\Phi(a)p^{-d}$ and $g_1(q/p)^{-d}$ are
trace class by Lemmas 2.3 and 2.4(i), and
3) product of a bounded operator and a trace class
operator is trace class.

\qed

In fact, we have proved the following estimate on matrix coefficients
of intertwiners.

\proclaim{Lemma 3.2}
Let $\Phi\in I_{\l \nu k}(V^*)$.
Let $||\cdot ||$ be any norm on $V^*$. Then for any $q\in\C^*$ such that
$|q|<1$ there exists a constant $C_q$ such that for any
homogeneus vectors $v\in L_{\nu,k},w\in L_{\l,k}$ we have
$$
|| (v,\Phi w)||\le C_q|z_n|^{d(w)}|z_1|^{-d(v)}q^{d(v)+d(w)},
$$
where $d(v),d(w)$ are the degrees of $v,w$.
\endproclaim

{\bf 3.2. Conjugacy classes in $\tilde G_q$ and holomorphic principal bundles.}

Before we start proving the other results of Chapter 2, we have to
describe the conjugacy classes of the action of $\hat G$ on $\tilde G_q$.
First we describe conjugacy classes of the action of $LG$
on $\check G_q=\{(q,g),g\in LG\}$.

Let $(q,g)\in\check G_q$. Consider the elliptic curve
$E_q=\C^*/q^{\Bbb Z}$. Let $B_g$ be the holomorphic principal $G$-bundle
over $E_q$ defined as follows. We view $E_q$ as the annulus
$|q|\le |z|\le 1$ in the complex plane, whose boundaries are identified
with each other via the map $z\to qz$.
We start with a trivial bundle on the annulus,
and then define the bundle $B_g$ on $E_q$ by describing the attachment map
of fibers over the points identified under $z\to qz$:
$f(qz)=g(z)f(z)$ ($f$ takes values in $G$).

Then we have:

\proclaim{Theorem 3.3} (E.Looijenga)

(i) Two elements $(q,g_1),
(q,g_2)\in\check G_q$
are $LG$-conjugate to each other if and only if the corresponding
holomorphic principal bundles $B_{g_1}$ and $B_{g_2}$ are isomorphic.

(ii) For any principal $G$-bundle $B$ over $E_q$ there exists
an element $g\in LG$ such that $B=B_g$. Thus, $LG$-conjugacy classes
in $\check G_q$ are in one-to-one correspondence with
 holomorphic principal $G$-bundles
over $E_q$.
\endproclaim

The proof of this theorem is straightforward (see, e.g.,\cite{EF})

Principal bundles over an elliptic curve can be classified.
However, we will not need a complete classification.
All we will need is classification of flat and unitary bundles.

\proclaim{Definition 3.1} A holomorphic principal $G$-bundle $B$
over $E_q$ is called
flat and unitary if it admits a flat and unitary connection
(i.e. a flat connection with monodromy in $K$) compatible to the complex
structure.
\endproclaim

\proclaim{Theorem 3.4} (\cite{NS},\cite{R}) In any holomorphic family $\Cal T$
of holomorphic principal $G$-bundles over $E_q$ almost every bundle is
flat and unitary
(i.e. this is true on a nonempty Zariski open subset of $\Cal T$).
\endproclaim

\proclaim{Definition 3.2} We say that $(q,g)\in \check G_q$
is semisimple if the bundle $B_g$ is flat and unitary.
\endproclaim

\proclaim{Corollary 3.5}
Almost all elements of $\check G_q$ are semisimple.
That is, the set of elements $g\in LG$
such that $(q,g)$ is semisimple
contains a nonempty Zariski open set.
\endproclaim

Conjugacy classes of semisimple elements are very easy to classify.
Namely, from the definition of semisimplicity one gets

\proclaim{Proposition 3.6} Every semisimple element in $\check G_q$
is $LG$-conjugate to an element of the form $(q,\xi )$, where
$\xi\in H$. Two elements $(q,\xi _1),(q,\xi _2)$ are conjugate iff
$\xi_1,\xi_2$ are in the same orbit of the affine Weyl group $\hat W$
on $H$.
\endproclaim

Thus, the set of conjugacy classes of semisimple elements in $\check G_q$
is isomorphic to $H/\hat W$.

Let $Q^{\vee}$ be the dual root lattice of $G$, spanned (over $\Z$) by the
elements $\frac{2\alpha}{\<\alpha,\alpha\> }$, $\alpha\in  R^+$.
Let $A_q(G)$ be the abelian variety $H/q^{Q^{\vee}}$ (it is isomorphic
to the product of $r$ copies of the elliptic curve
 $E_q$, where $r$ is the rank of $G$). There is a natural action
of the Weyl group $W$ on $A_q(G)$, and, as we have seen, the moduli space
of flat and unitary bundles (=semisimple conjugacy classes in $\check G_q$)
is identified with $A_q(G)/W$.

Let us now consider semisimple $\hat G$-conjugacy classes in $\tilde G_q$,
i.e. those that project to semisimple $LG$-conjugacy classes of $\check G_q$.
Since $\tilde G_q$ is a fiber bundle over $\check G_q$ with fiber $\C^*$, and
conjugation by an element of $\hat G$ is an automorphism of this bundle,
the set of semisimple $\hat G$-conjugacy classes in $\tilde G_q$ has to be
the total space of a $\C^*$-bundle over the set of semisimple $LG$-conjugacy
classes in $\check G_q$. The exact type of this bundle is determined
as follows.

Let $\tilde H_q$ be the set of all elements in $\tilde G_q$ that
project to an element of the form $(q,\xi )\in\check G_q$, $\xi\in H$.
As a complex manifold, it is isomorphic to $H\times \C^*$.
The set of semisimple $\hat G$-conjugacy classes in $\tilde G_q$
is the quotient $\tilde H_q/\hat W$, where $\hat W$ acts
as follows:
$$
\gather
\hat W=W\ltimes Q^{\vee}=\{(w,\beta),w\in W,\beta\in Q^{\vee}\},\\
 (w,1)\circ(q,\xi ,u)=(q,w(\xi),u),  w\in W, \\
(1,\beta)\circ(q,\xi ,u)=
(q,\xi e^{-2\pi i\beta},uq^{\frac{1}{2}\<\beta,\beta\> }\beta(\xi^{-1})),
\tag 3.3\endgather
$$
where $\beta(\xi^{-1})$ denotes the value of $\beta$, as a character of $H$,
at the point $\xi^{-1}$. This shows that the space $\tilde H_q/Q^{\vee}$,
as a complex manifold, is isomorphic to the set of nonzero vectors in
the total space of a certain holomorphic line bundle $\Cal L$ over
$A_q(G)$, defined by (3.3). This bundle was introduced by Looijenga
(\cite{Lo}). It is invariant under the natural action of the Weyl group $W$.

Now, let $V^*$ be a finite-dimensional
representation of $\hat G$. Then $V^*[0]$ is naturally
a representation of the affine Weyl group $\hat W$.
In particular, the action in $V^*[0]$ of the subgroup $Q^{\vee}$ in $\hat W$
naturally defines a flat vector bundle over $A_q(G)$ with fiber
$V^*[0]$ (the total space of this bundle is $\tilde H_q\times V^*[0]/Q^{\vee}$,
where $Q^{\vee}$ acts diagonally). We will denote it by $\Cal B_V$.
Again, it is invariant under a natural action of $W$, which gives rise
to an action of $W$ on its sections.

Further, let $\Psi: \tilde G_q\to V^*$ be an equivariant function
of degree $k$. Consider its restriction to $\tilde H_q$.
By Corollary 3.5, this restriction
uniquely determines $\Psi$. On the other hand, this restriction has
to be $\hat W$-equivariant, which means that it defines
a $W$-invariant section of the vector bundle $\Cal L^k\otimes \Cal B_V$
(this follows from formulas (3.3)) Thus, we
have obtained the following statement.

\proclaim{Proposition 3.7} Restriction of equivariant
functions to $\tilde H_q$ defines an embedding of vector spaces
$F_k^V(\tilde G_q)\to \Gamma^W(\Cal L^k\otimes \Cal B_V)$, where
$\Gamma^W(\Cal L^k\otimes \Cal B_V)$ denotes the space of $W$-invariant global
sections of the holomorphic vector bundle $\Cal L^k\otimes \Cal B_V$.
\endproclaim

In particular, this proposition implies that the space of
equivariant functions $F_k^V(\tilde G_q)$ is finite-dimensional.

{\bf Remark.} This embedding is not, in general, an isomorphism.
It is an isomorphism if and only if $V$ is
a trivial representation. Below we will formulate the necessary and
sufficient condition for a section of $\Cal L^k\otimes \Cal B_V$ to extend
to an equivariant function, i.e. to be in the image of this embedding.

\vskip .1in

{\bf 3.3. Proof of Theorem 2.6.}

Let $\tilde H$ denote the Cartan subgroup of $\tilde G$.
Elements of $\tilde H$ will be denoted as $(q,\xi ,u)$, $q,u\in\C^*$
$\xi\in H$ (as before, $u$ represents the central component).
By $\tilde H_{q_0}$ we denote the set of elements
$(q,\xi ,u)\in\tilde H$
with $q=q_0$. Let $\tilde \frak h$ be the Lie algebra of $\tilde H$.

Let ${\hat\l}\in \hat P_k^+$, $W_{\hat\l}$ be the stabilizer of
${\hat\l}$ in the affine Weyl group, $v\in V^*[0]^{W_{\hat\l}}$,
where $v\in V^*[0]^{W_{\hat\l}}$ denotes the space of vectors in
$V^*[0]$
stable under $W_{\hat\l}$.
Introduce the ``orbitsum'' $m_{\hat\l}(v)$ --
the theta-function
$$
m_{\hat\l}(v)(q,\xi ,u)=\sum_{w\in\hat W}e^{2\pi i\<w^{-1}{\hat\l},x\> }wv,
x\in \tilde{\frak h}, (q,\xi ,u)=e^{2\pi ix}\in \tilde H.\tag 3.4
$$
Obviously, this function linearly depends on $v$. Also,
observe that $m_{\hat\l+Nc^*}(v)=q^Nm_{\hat\l}(v)$.

Let $\hat P^+_{k0}=\{{\hat\l}\in \hat P^+_k: \<{\hat\l},d\> =0\}$.
The following statements are checked directly.

\proclaim{Lemma 3.8}
The restriction of $m_{\hat\l}(v)$ to $\tilde H_q$ is
a $\hat W$-equivariant function, i.e. it defines
a regular $W$-invariant
global section of the vector bundle $\Cal L^k\otimes \Cal B_V$.
\endproclaim

\proclaim{Lemma 3.9} Let $v_1^{\hat\l},
...,v_{s(\hat\l)}^{\hat\l}$ be a basis of the space $V^*[0]^{W_{\hat\l}}$.
Then the functions
$m_{\hat\l}(v_i)$, ${\hat\l}\in \hat P^+_{k0}$, $i=1,...,s(\hat\l)$,
are a basis
of the space $\Gamma^W(\Cal L^k\otimes \Cal B_V)$ of $\hat W$-equivariant
functions on $\tilde H_q$. Thus, the space $\Gamma^W(\Cal L^k\otimes \Cal B_V)$
is isomorphic to $\oplus_{\hat\l\in\hat P_{k0}^+}V^*[0]^{W_{\hat\l}}$.
\endproclaim

These statements immediately imply statement (i) of Theorem 2.6, since
the set $\hat P_k^+$ is empty for $k<0$.

Let us now prove statements (ii) and (iii) of Theorem 2.6.
Assume that $k\ge 0$.
Let $\Psi: \tilde G_q\to V^*$ be an equivariant function of degree $k$.
Let us restrict this function to $\tilde H_q=H\times \C^*$,
and express this restriction as a linear combination
of orbitsums:
$$
\Psi(q,\xi ,u)=\sum_{{\hat\l}\in \hat P_{k0}^+}
m_{\hat\l}(v_{{\hat\l}})(q,\xi ,u), v_{\hat\l}\in V^*[0]^{W_{\hat\l}}.\tag 3.5
$$
Now we will deduce a necessary condition for a function
$\Psi\in \Gamma^W(\Cal L^k\otimes \Cal B_V)$ to extend to an equivariant
function on $\tilde G_q$. This condition will later turn out to be also
sufficient. It plays the central role in the proof.

\proclaim{Lemma 3.10} If $\Psi\in \Gamma^W(\Cal L^k\otimes \Cal B_V)$
extends to an equivariant function on $\tilde G_q$ then
for every positive integer $m$, every
positive root $\alpha\in \hat R$ and any root element
$F\in\hat \g$ such that $[h,F]=\alpha(h)F$, $h\in\tilde {\frak h}$
the function $F^m\Psi(q,\xi ,u)$ is divisible
(in the ring of holomorphic functions of $\xi,u$) by the
function $(1-e^{2\pi i\<\alpha,x\> })^m$, where (as before)
$x\in\tilde{\frak h}$ is such that $e^{2\pi ix}=(q,\xi ,u)$.
\endproclaim

{\bf Remark. } Note that the condition on $\Psi$ in Lemma 3.10 is
equivalent to condition (4.10) in \cite{FG} for Chern-Simons states.

\demo{Proof} Let $y$ be a variable. Let $\Psi$ be an equivariant function.
The equivariance condition tells us the following:
$$
e^{-yF}\Psi(e^{yF}e^{2\pi ix}e^{-yF})=\Psi(e^{2\pi ix}).\tag 3.6
$$
Let $\eta=e^{2\pi i\<x,\alpha\> }$. Then $e^{yF}e^{2\pi ix}e^{-yF}=
e^{2\pi ix}e^{y(\eta-1)F}$. Thus
$$
e^{\frac{yF}{\eta-1}}\Psi(e^{2\pi ix})=\Psi(e^{2\pi ix}e^{yF}).\tag 3.7
$$
Comparing the Taylor expansions in powers of $y$ of both sides of (3.7), we get
$$
\frac{F^m\Psi(e^{2\pi ix})}{(\eta-1)^m}=L_F^m\Psi(e^{2\pi ix})\tag 3.8
$$
where $L_F$ denotes the operator of Lie derivative along the left-invariant
vector field on $\tilde G$ whose value at $1$ is $F\in\tilde \g$.
Since the right hand side of identity
(3.8) is holomorphic, this identity implies the statement of the Lemma.
\enddemo

Let $U_q\subset \oplus_{\hat\l\in\hat P^+_{k0}}
V^*[0]^{W_{\hat\l}}$ be the subspace
of all functions satisfying the property from Lemma 3.10.
(Lemma 3.10 states that the restriction of an equivariant
function belongs to $U_q$). The space $U_q$ is given by a system
of linear equations expressing divisibility of $F^m\Psi$ by
$(1-e^{2\pi i\<\alpha,x\> })^m$. The coefficients of these equations
are meromorphic near $q=0$, which implies
that it is possible to find a basis of $U_q$,
$B_1(q),...,B_M(q)$, such that $B_i(q)$ are holomorphic
in $0<|q|<Q'$ for some $Q'=Q'(V)$ and meromorphic at $q=0$
(the argument in Section 5
of \cite{FG} shows that in fact $Q'(V)=Q(V)$).
If $0<|q_0|<Q'$, and
$\Psi$ is any equivariant function on $\tilde G_{q_0}$ then
$\Psi|_{\tilde H_{q_0}}=\sum_{j=1}^MC_jB_j(q_0)$.
Let us now extend the function $\Psi$ to arbitrary values of $q$
by setting $\Psi|_{\tilde H_q}=\sum_{j=1}^MC_jB_j(q)$.
Then the function $\Psi$ is holomorphic for $0<|q|<Q'$ and
meromorphic at $q=0$. This means that the function $\Psi$ is
representable in the form
$$
\Psi(q,\xi ,u)=\sum_{{\hat\l}\in \hat P_{k}^+,\hat\l\le\hat\l_0}
m_{\hat\l}(v_{{\hat\l}})(q,\xi ,u), v_{\hat\l}\in V^*[0]^{W_{\hat\l}},
$$
for some $\hat\l_0\in \hat P_{k}^+$,
and this series is convergent for $0<|q|<Q'$ and any $\xi,u$.
Our purpose is to show that this implies that
$\Psi$ is a linear combination of traces of intertwiners.

\proclaim{Lemma 3.11} Assume that
$$
\Psi(q,\xi ,u)=\sum_{{\hat\l}\in \hat P_{k}^+,\hat\l\le\hat\l_0}
m_{\hat\l}(v_{{\hat\l}})(q,\xi ,u), v_{\hat\l}\in V^*[0]^{W_{\hat\l}},\tag 3.9
$$
and this series is convergent in $0<|q|<Q'$. Suppose that also
$\Psi\in U_q$ for all $q$ such that $0<|q|<Q'$.
Then $\Psi$ is a linear combination of functions of the form (2.1)
with coefficients in $\C((q))$ which converge in $0<|q|<Q'$.
\endproclaim

\demo{Proof}
Let $X$ be the weight diagram of $\Psi$, i.e. the set of all weights
from $\hat P$
which occur in the Laurent expansion of $\Psi$ in $q,\xi ,u$, and all
lower weights.
Clearly, $X$ is $\hat W$-invariant. Let $X^+$ be the set of positive integral
weights in $X$ ($X=\hat WX^+$)
In order to prove the Lemma, it is enough to prove
that one can subtract a multiple of a function of the form (2.1)
from $\Psi$ so that the weight diagram of the
obtained function will be a proper subset of $X$.

A weight $\hat\l\in X^+$ is called extremal if
for every $\hat\mu\in \hat P^+_k$ such that $\hat\mu>\hat\l$
one has $\hat\mu\notin X$.
Clearly, extremal weights exist: $X$ is bounded from above because
$\Psi$ is meromorphic in $q$ at $q=0$.

We write the function $\Psi$ in the form
$$
\Psi(q,\xi ,u)=\sum_{\hat\l\in X^+}m_{\hat\l}(v_{\hat\l})(q,\xi ,u).\tag 3.10
$$
Let $\hat\l\in X^+$ be an extremal weight. Consider the vector $v_{\hat\l}$.

\proclaim{ Proposition 3.12 }
For every positive simple root $\alpha_j$ of $\hat\g$
($i=0,...,r$) one has $E_{\alpha_j}^{\<\hat\l,\alpha_j^{\vee}\> +1}
v_{\hat\l}=0$,
where $E_{\alpha_j}$ are the root elements corresponding to $\alpha_j$, and
$\alpha_j^{\vee}=\frac{2\alpha_j}{\<\alpha_j,\alpha_j\> }$.
\endproclaim

{\it Proof of Proposition 3.12. }  We know from Lemma 3.10 that
$E_{\alpha_j}^{\<\hat\l,\alpha_j^{\vee}\> +1}\Psi$ is divisible by
$(1-e^{2\pi i\<\alpha_j,x\> })^{\<\hat\l,\alpha_j^{\vee}\> +1}$. This implies
that the same is true for the sum $S$ of all terms of the Fourier
expansion of $E_{\alpha_j}^{\<\hat\l,\alpha_j^{\vee}\> +1}\Psi$
whose weights lie on the line $\hat\mu=\hat\l+\tau\alpha_j$, $\tau\in\Bbb R$.
But because of 1) the $\hat W$-symmetry, and 2) the extremality of $\hat\l$,
weights of all terms in the sum $S$ have to be between
$\hat\l$ and $s_{\alpha_j}(\hat\l)=\hat\l-\<\hat\l,\alpha_j^{\vee}\> \alpha_j$
(where $s_{\alpha}$ is the simple reflection corresponding to the root
$\alpha$). Thus, $S$ has the form $S(x)=e^{2\pi i\<\hat\l,x\> }S_0(x)$, where
$S_0(x)$ is a polynomial in $\eta=e^{-2\pi i\<\alpha_j,x\>} $
of degree $\<\hat\l,\alpha_j^{\vee}\> $. But we know that
$S_0(x)$ is divisible by
$(1-e^{2\pi i\<\alpha_j,x\> })^{\<\hat\l,\alpha_j^{\vee}\> +1}$.
This implies that $S_0(x)=0$, i.e. $S(x)=0$, which implies
Proposition 3.12. \qed

We will also need a standard fact from representation theory.

\vskip .03in

Let $L_{\hat\l}$ denote the integrable highest weight $\tilde\g_{pol}$-module
with highest weight $\hat\l$. If $\hat\l= \l+kd^*+Nc^*$
($\l\in\frak h^*$) then $L_{\hat\l}$ is isomorphic to $L_{\l,k}$ as a
$\hat\g_{pol}$-module, but the action of $d$ is defined by the condition that
its eigenvalue on the vacuum vector is $N$.

\proclaim{Proposition 3.13. } (cf. \cite{TK})
 Let $v\in V^*[0]$. An intertwining operator
for $\hat \g_{pol}$, $\Phi: L_{\hat\l}\to \hat L_{\hat \l}\otimes V^*$
with highest matrix coefficient $(vac, \Phi\cdot vac)=v$
($vac$ denotes the vacuum vector) exists if and only if
$E_{\alpha_j}^{\<\hat\l,\alpha_j^{\vee}\> +1}v=0$, $j=0,...,r$.
If such an operator exists, it is unique.
\endproclaim

\vskip .03in

Now let us finish the proof of Lemma 3.11.
 According to Propositions 3.12, 3.13,
for any extremal weight $\hat\l\in X^+$ there exists
an operator $\Phi: L_{\hat\l}\to\hat L_{\hat\l}\otimes V^*$
with the property $(vac,\Phi vac)=v_{\hat\l}$.
Let us construct from it an equivariant function $\Psi_{\hat\l}$
using formula (2.1). Now consider the function
$\Psi'=\Psi-\Psi_{\hat\l}$. Let $X'$ be the weight diagram of
$\Psi'$. Clearly, $X'\subset X$. Moreover, $X'$ is a proper subset in
$X$, since $\hat\l\notin X'$ (the $\hat\l$-coefficients
in $\Psi$ and $\Psi_{\hat\l}$ were the same by construction), so they
cancelled each other.

Now take any extremal weight in $X'$ and apply the above procedure
to it, and so on. Continuing this to infinity, we will obtain
an expansion of $\Psi$ in a series of functions of the form (2.1).
After we use the fact that $\Tr|_{L_{\hat\l}}(\Phi(a) q^{-d}g)=q^{-N}
\Tr|_{L_{\l,k}}(\Phi(a) q^{-d}g)$ if $\hat\l=\l+kd^*+Nc^*$,
we obtain a representation
of $\Psi$ as a finite linear combination
of functions of the form $\Tr|_{L_{\l,k}}(\Phi(a)q^{-d}g)$
whose coefficients are Laurent series in $q$. It is obvious
that these series define holomorphic functions of $q$ in $0<|q|<Q'$
(because $\Psi$ is holomorphic in this region). Lemma 3.11 is proved.
\enddemo

Now we finish the proof of statements (ii) and (iii)
of Theorem 2.6. We have shown that any equivariant function is a
(finite) linear combination of functions of the form (2.1).

The facts that 1) the subspaces $S(V,L_{\l,k})$ are disjoint, and 2) that
linearly independent intertwiners $\Phi$ define (by (2.1)) linearly independent
functions $\Psi$, follow from the fact that after restriction to the
torus these functions are eigenfunctions
the parabolic operator (2.5).
Indeed, assume
that at some
special value of $q$ there is a nontrivial linear relation between
the functions $\Psi$. In particular, it holds on
$\tilde H_q$. But the restriction of $\Psi$ on $\tilde H_q$ satisfies
the parabolic differential equation: after rescaling by a power of $q$
it is annihilated by operator (2.5). Therefore, using the parabolic
equation as a connection, we can transport our linear relation between
$\Psi$'s to other values of
$q$. In particular, we can take the limit $q\to 0$. But in this limit it
is obvious that any linear relation between $\Psi$'s is trivial, so we
get a contradiction.

This implies
Theorem 2.6(ii),(iii).

\qed

{\bf 3.4. Proof of Theorems 2.8, 2.14.}

{\it Proof of Theorem 2.8. }

(i) Let us first show admissibility of holomorphic equivariant functions.
By Theorem 2.6, every equivariant function is a linear combination
of traces of the form (2.1) with coefficients, holomorphically depending on
$q$. Also, it is obvious that if $\Psi$ is an admissible function,
and $\phi(q)$ is holomorphic, then $\phi(q)\Psi$ is also admissible
(indeed, all terms in the sum (2.2) except $2(c+h^{\vee})d$ commute with
multiplication by $\phi$). Therefore, it is enough to show
that all functions of the form (2.1) are admissible.
This is done as follows.

We have
$$
\Delta_{\tilde G}^{(N)}\Tr|_{L_{\l,k}}(\Phi(a)q^{-d}g)=
\Tr|_{L_{\l,k}}(\Phi(a)q^{-d}g
(2(c+h^{\vee})d+\sum_{a\in B}\sum_{n=-N}^N:a[n]a[-n]:)).\tag
3.11
$$
Therefore, to prove the existence of the limit
of (3.11) as $n\to +\infty$, it is enough to check that
$$
\Tr|_{L_{\l,k}}(\Phi(a)q^{-d}g
:a[n]a[-n]:)=o(P^n),\ n\to +\infty,\tag 3.12
$$
for some $P<1$.

Choose $p\in\C$ such that $1>|p|>|q|$ and
$|p/z_n|<|a|<|1/z_1|$. Set $A=\Phi(a)p^{-d}g_1$,
$g_1=(q/p)^dg(q/p)^{-d}\in \hat G$, and $B_n=:a[n]a[-n]:(q/p)^{-d}$.
Then
$$AB_n=\Phi(a)q^{-d}g
:a[n]a[-n]:$$

We know that $A$ is bounded (Lemma 3.4(ii)). Therefore,
to prove (3.12), it is enough to show that $||B_n||=o(P^n)$.
This statement can be checked directly: the proof is based on the fact that
the operator $B_n$ kills all homogeneous vectors of degree $>-n$, and
the estimate in Lemma 3.2 in \cite{GW1}, which shows that
the matrix elements of the operator representing an element
of $U(\hat\g_{pol})$ in $L_{\l,k}$ grow slower than exponentially
with degree.

Let now $\Psi$ be a meromorphic equivariant function of degree
$l\in\Z$. Then we can write $\Psi$ in the form $\Psi=\Psi_0/\chi$, where
$\Psi_0$ is a holomorphic equivariant function of degree $k$, and
$\chi$ is a central function of degree $k-l$ ($k\ge l$)
Applying $\Delta_{\tilde G}^{(N)}$ to the ratio $\Psi_0/\chi$,
we see that in order to establish the convergence
of truncated sums $(\Delta_{\tilde G})^{(N)}
\Psi$ to some limit, it is enough to show
that if $\Psi$ is a holomorphic
 equivariant function of a positive degree then 1) $L_{a[-n]}\Psi=O(P^n)$,
$n\to +\infty$, for some $P<1$, and 2) $L_{a[n]}\Psi=O(R^n)$, $n\to+\infty$,
 for every $R>1$ (here $L_x$ denotes the Lie derivative along the
left-invariant vector field on $\tilde G$ defined by $x\in\tilde\g$).
By Theorem 2.6, it is enough to do it for functions of the form (2.1).

 From (2.1) we have $L_{a[n]}\Psi=\Tr|_{L_{\l,k}}(\Phi(a)q^{-d}ga[n])=
\Tr(AD_n)$, where $A=\Phi(a)p^{-d}g$, $D_n=(q/p)^{-d}a_n$,
and $p$ is chosen as above. Further, we have
$\Tr(AD_n)=O(||A||\cdot||D_n||)$.
Thus, it suffices to prove
that 1) $||D_{-n}||=O(P^n)$,
$n\to +\infty$, for some $P<1$, and 2) $||D_n||=O(R^n)$, $n\to+\infty$,
 for every $R>1$. This statement is again checked directly:
it follows from Lemma 3.2 in \cite{GW1}.

(iv) It follows from the definition of the Laplacian that
$$
\Delta_{\tilde G} \Tr(\Phi(a)q^{-d}g)=\Tr(\Phi(a)q^{-d}gC),\tag 3.13
$$
where $C$ is the Casimir-Sugawara element. Since
$C|_{L_{\l,k}}=\<\l,\l+2\rho\> $, we get the desired formula.

(ii) Let $\Psi$ be an equivariant function. We must show that
so is $\Delta_{\tilde G}\Psi$. Let $x\in\tilde \g_{pol}$. Since
$\Delta_{\tilde G}$ was constructed from a central element,
$[L_x,\Delta_{\tilde G}]=0$. Therefore, $L_x\Delta_{\tilde G}\Psi=
\Delta_{\tilde G} L_x\Psi=\Delta_{\tilde G} (\pi_{V^*}(x)\Psi)=
\pi_{V^*}(x)\Delta_{\tilde G}\Psi$, i.e. $\Delta_{\tilde G}
\Psi$ is equivariant, Q.E.D.

(iii) is obvious, since the action of $\hat G$ on $F_k^V$ by conjugacy
coincides with its action on values (by conjugacy invariance), so it
commutes with $\Delta_{\tilde G}$.

{\it Proof of Theorem 2.14}

(i) The proof is analogous to the proof of Statement (i) in Theorem 2.8.

(ii) Let $\Psi$ be an equivariant function. We must show that
so is $\Delta_i\Psi$. Let $x\in\tilde \g_{pol}$. Since
$\Delta_i$ were constructed from central elements,
$[L_x,\Delta_i]=0$. Therefore, $L_x\Delta_i\Psi=
\Delta_i L_x\Psi=\Delta_i (\pi_{V^*}(x)\Psi)=
\pi_{V^*}(x)\Delta_i\Psi$, i.e. $\Delta_i\Psi$ is equivariant, Q.E.D.

(iii) follows form the relation $[L_x,\Delta_i]=0$.

\end